\documentclass[preprint,showpacs,preprintnumbers,amsmath,amssymb]{revtex4}
\usepackage{amsmath}

\usepackage{graphicx}
\usepackage{pdfpages}
\usepackage{dcolumn}
\usepackage{bm}
\usepackage{amsmath} 
\usepackage{amsfonts}
\usepackage{amssymb}
\usepackage{url}
\usepackage{microtype}
\usepackage{graphicx}%

\newcommand{\qed}{\nobreak \ifvmode \relax \else
      \ifdim\lastskip<1.5em \hskip-\lastskip
      \hskip1.5em plus0em minus0.5em \fi \nobreak
      \vrule height0.75em width0.5em depth0.25em\fi}

\begin{document}

\preprint{}
\title{Jagged Islands of Bound Entanglement and Witness-Parameterized Probabilities}
\author{Paul B. Slater}
 \email{slater@kitp.ucsb.edu}
\affiliation{%
Kavli Institute for Theoretical Physics, University of California, Santa Barbara, CA 93106-4030\\
}
\date{\today}
            
\begin{abstract}
We report several witness-parameterized families of bound-entangled probabilities. Two pertain to the $d=3$ (two-qutrit)  and a third to the $d=4$ (two-ququart) subsets analyzed by Hiesmayr and L{\"o}ffler  of ``magic" simplices of Bell states that were introduced by Baumgartner, Hiesmayr and Narnhofer.  The Hilbert-Schmidt probabilities of positive-partial-transpose (PPT)  states--within which we search for bound-entangled states--are 
$\frac{8 \pi }{27 \sqrt{3}} \approx 0.537422$ ($d=3$) and $\frac{1}{2}+\frac{\log \left(2-\sqrt{3}\right)}{8 \sqrt{3}} \approx 0.404957$ ($d=4$). We obtain bound-entangled probabilities  of   $-\frac{4}{9}+\frac{4 \pi }{27 \sqrt{3}}+\frac{\log (3)}{6} \approx 0.00736862$ and $\frac{-204+7 \log (7)+168 \sqrt{3} \cos ^{-1}\left(\frac{11}{14}\right)}{1134}  \approx 0.00325613$ ($d=3$) and $\frac{8 \log (2)}{27}-\frac{59}{288} \approx 0.00051583$ and $\frac{24 \text{csch}^{-1}\left(\frac{8}{\sqrt{17}}\right)}{17 \sqrt{17}}-\frac{91}{544} \approx 0.00218722$ ($d=4$). (For $d=3$, we also obtain $\frac{2}{81} \left(4 \sqrt{3} \pi -21\right) \approx 0.0189035$ based on the realignment criterion. 
Thus, the total entanglement probability appears to equal $(1-\frac{8 \pi }{27 \sqrt{3}})+\frac{2}{81} \left(4 \sqrt{3} \pi -21\right) = \frac{13}{27} \approx 0.481481$.) The families, encompassing these results, are parameterized using generalized Choi  and Jafarizadeh-Behzadi-Akbari witnesses. In the $d=3$, analyses, we  utilized the  mutually unbiased bases (MUB) test of Hiesmayr and L{\"o}ffler, and also the Choi $W^{(+)}$ test. 
The same bound-entangled probability was achieved with both--the sets (``jagged islands") detected having void intersection. The  entanglement (bound and ``non-bound"/``free") probability for  both was $\frac{1}{6} \approx 0.16667$,  while their union and intersection gave $\frac{2}{9} \approx 0.22222$ and  $\frac{1}{9} \approx 0.11111$.  Further, we examine generalized Horodecki states, as well as estimating  PPT-probabilities of approximately 0.39339 (very well-fitted by  $\frac{7 \pi}{25 \sqrt{5}} \approx 0.39338962$) and 0.115732 (conjecturally, $\frac{1}{8}+\frac{\log \left(3-\sqrt{5}\right)}{13 \sqrt{5}} \approx 0.115737$) for the original (8- [two-qutrit] and 15 [two-ququart]-dimensional) magic simplices themselves.
\end{abstract}
 

\maketitle
\section{Introduction}
In their landmark 1998 paper, ``Volume of the set of separable states",  {\.Z}yczkowski, Horodecki, Sanpera and Lewenstein stated: ``As it was mentioned in the introduction for $N \ge 8$ there are states which
are inseparable but have positive partial transposition.
Moreover, it has been recently shown that all states of such type
represent `bound' entanglement in the sense that they cannot
be distilled to the singlet form.
The immediate question that arises is how frequently such peculiar
states appear in the set of all the states of a given composite
system. This question is related to the role of time reversal in 
the context of entanglement of mixed states....we provide a qualitative argument that the volume of the set of those states
 is also nonzero" \cite[sec. V]{zyczkowski1998volume}. 
 
 We will here offer some {\it quantitative} arguments in this direction (cf. \cite{bae2009detection}), where the sets of primary interest are the ``magic" simplices of Bell states (sec.~\ref{MagicSimplicesAnalyses}) \cite{baumgartner2006state,baumgartner2008geometry,derkacz2007entanglement,sentis2018bound}--for which it had been noted that the ``Hilbert-Schmidt metric defines a natural metric on the space state" \cite{baumgartner2006state}--and generalized Horodecki states (sec.~\ref{GeneralizedHorodecki}) \cite{chruscinski2011family,jafarizadeh2009entanglement}.
\section{Magic Simplices Analyses} \label{MagicSimplicesAnalyses}
Within the magic simplex setting of Baumgartner, Hiesmayr and Narnhofer \cite{baumgartner2006state,baumgartner2008geometry}, the case of bound entanglement of two photonic qutrits  using the orbital angular momentum degree of freedom of light was investigated in a 2013 paper  of B. C. Hiesmayr and W. L{\"o}ffler \cite{hiesmayr2013complementarity}. They noted that this was the simplest case of bound entanglement, with ``complications, such as those arising in multipartite systems, not occurring''.

Their equation (7)  for a density matrix $\rho_d$ in the ($d^2-1$)-dimensional  simplices  ($d =3, 4$) took the form,
\begin{equation} \label{densitymatrix}
\rho_d= \frac{q_4 (1-\delta (d-3)) \sum _{z=2}^{d-2} \left(\sum _{i=0}^{d-1}
   P_{i,z}\right)}{d}+\frac{q_2 \sum _{i=1}^{d-1} P_{i,0}}{(d-1) (d+1)}+\frac{q_3 \sum
   _{i=0}^{d-1} P_{i,1}}{d}+
\end{equation}
\begin{displaymath}
   \frac{\left(-\frac{q_1}{d^2-d-1}-\frac{q_2}{d+1}-(d-3)
   q_4-q_3+1\right) \text{IdentityMatrix}\left[d^2\right]}{d^2}+\frac{q_1
   P_{0,0}}{d^2-d-1},
\end{displaymath}
the $P_{i,j}$'s being orthonormal Bell states. (No explicit ranges were given for the $q$'s, and  our initial analyses assumed that they would have to be nonnegative. But, it, then, seemed somewhat puzzling that a main example of Hiesmayr and L{\"o}ffler employed negative $q$'s. Also, we observed that bound entanglement did not seem possible with strictly nonnegative $q$'s. Eventually, we arrived at the clearly powerful change-of-variables approach--to be shortly detailed--greatly facilitating the exact integrations we had been attempting.)

``This family also includes for $d=3$ the one-parameter Horodecki--state,
the first found bound entangled state. Namely,
for $q_{1}=\frac{30-5\lambda}{21},\; q_{2}=-\frac{8\lambda}{21},\; q_{3}=\frac{5-2\lambda}{7}$
with $\lambda\in[0,5]$. This state is PPT for $\lambda\in[1,4]$
and was shown to be bound entangled for $\lambda\in\{3,4]$" \cite{hiesmayr2013complementarity}. Let us note now--as a prototypical example of our subsequent more demanding three- and four-parameter calculations--that the Hilbert-Schmidt PPT-probability for this one-parameter ($\lambda$) Horodecki-state is $\frac{3}{5}$, the probability of entanglement is also $\frac{3}{5}$, and the bound-entangled probability, $\frac{1}{5}$ (cf. \cite{chruscinski2011family}, and sec.~\ref{GeneralizedHorodecki} below).
\subsection{Transformation between magic simplex parameters, and associated constraints} \label{Transformation}
In the $d=3$ (two-qutrit) framework, we transform between the nine nonnegative parameters  ($c[k,l] \geq 0$, with $k,l=0,\ldots2$), summing to 1, employed for the full/original magic simplex of Baumgartner, Hiesmayr and Narnhofer \cite[sec. 4]{baumgartner2008geometry}, and the three ($q_1,q_2, q_3$) of the Hiesmayr-L{\"o}fller subset. To do so, we use the three equations,
\begin{equation} \label{d=3Equality}
 c[0,0]=\frac{1}{180} (32 q_1-5 q_2-20 q_3+20) \equiv Q_1,
\end{equation} 
and
\begin{equation}
 c[k,l]=\frac{1}{180} 
 \left(-4 q_1-5 q_2+40 q_3+20\right) \equiv Q_2
\end{equation} 
 for $\{k,l\}={\{0,1\},\{0,2\},\{1,1\},\{1,2\},\{2,1\},\{2,2\}}$, and 
 \begin{equation}
   c[k,l]=\frac{1}{360} \left(-8 q_1+35 q_2-40 q_3+40\right)  \equiv Q_3
 \end{equation}
 for $\{k,l\}={\{1,0\},\{2,0\}}$. (For the indicated Horodecki-state, we have  $\{Q_1,Q_2,Q_3\}=\left\{-\frac{2 \lambda }{189}+\frac{2}{9 \lambda }+\frac{2}{7},\frac{1}{189} \left(-11
   \lambda +\frac{42}{\lambda }+45\right),\frac{\lambda ^2-21}{27 \lambda }\right\}.)$
 
\begin{figure}
    \centering
    \includegraphics{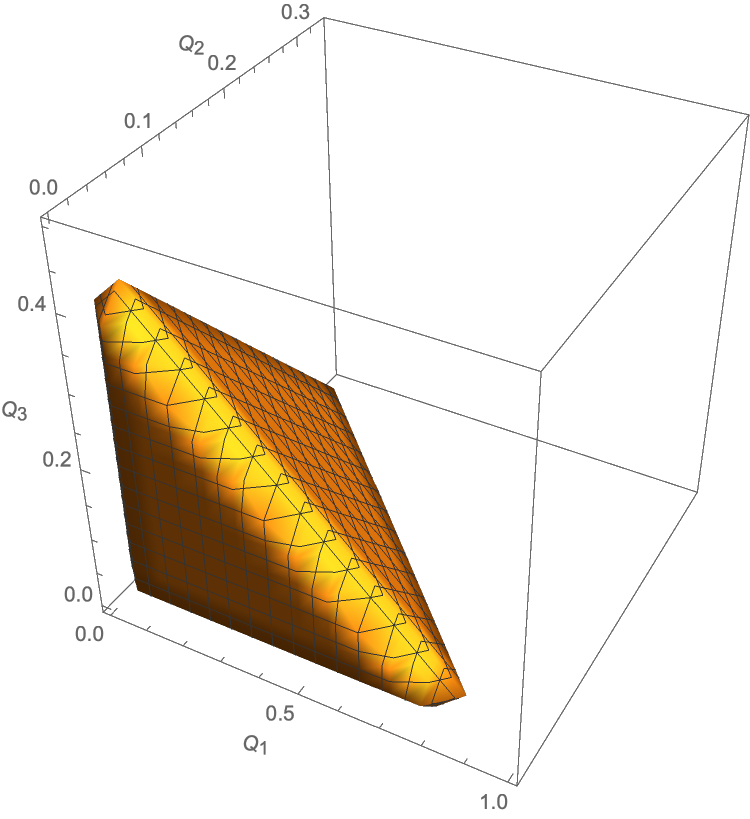}
    \caption{Convex set of Hiesmayr-L{\"o}ffler two-qutrit $d=3$ magic simplex density matrices $\rho_{3}$}
    \label{fig:ConvexTwoQutrits}
\end{figure}

If we, then, employ $Q_1, Q_2$ and $Q_3$ as our principal variables, rather than $q_1, q_2, q_3$, 
using the linear transformations,
\begin{equation}
\{q_1,q_2,q_3\} =\left\{\frac{5}{3} \left(4 Q_1+3 Q_2+2 Q_3-1\right),\frac{8}{3} \left(Q_1+3 Q_2+5
   Q_3-1\right),Q_1+6 Q_2+2 Q_3-1\right\},   
\end{equation}
our ensuing analyses simplify greatly. For example, the requirement that $\rho_3$ is a nonnegative definite density matrix--ensured by requiring that its nine leading nested minors all be nonnegative--is transformed from
\begin{displaymath}
4 q_1+5 q_2<40 q_3+20\land 10 q_3<4 q_1+5 q_2+10\land 4 q_1+5 q_2+20 q_3<20
\end{displaymath}
\begin{displaymath}
\land 5
   \left(q_2+4 q_3-4\right)<32 q_1\land \left(8 q_1-5 \left(7 q_2-8 q_3+8\right)\right)
   \left(8 q_1+q_2-8 q_3+8\right)<0 
\end{displaymath}
to 
\begin{equation}
Q_1>0\land Q_2>0\land Q_3>0\land Q_1+3 Q_2+2 Q_3<1.   
\end{equation}
In Fig.~\ref{fig:ConvexTwoQutrits} we show the convex set of possible $\rho_{3}$'s, in terms of this parameterization.
Additionally, the constraint that the partial transpose of $\rho_3$ is nonnegative definite becomes
\begin{equation} \label{d=3PPT}
 Q_1>0\land Q_3>0\land Q_1+3 Q_2+2 Q_3<1\land Q_1^2+3 Q_2 Q_1+\left(3
   Q_2+Q_3\right){}^2<3 Q_2+2 Q_1 Q_3.   
\end{equation}
(We report and employ the $d=4$ [two-ququart] analogues of the results (\ref{d=3Equality})-(\ref{d=3PPT}) in sec.~\ref{Twoququarts}.)
\subsubsection{MUB test}
Further, the Hiesmayr-L{\"o}ffler  mutually-unbiased-bases (MUB) criterion for bound entanglement, $I_4 = \Sigma_{k=1}^4 
C_{A_k,B_k} >2$, where $C_{A_k,B_k}$ are correlation functions for observables $A_k,B_k$ \cite[Fig. 1]{hiesmayr2013complementarity} takes the form
\begin{equation} \label{MUBconstraint}
 Q_1>3 Q_2+4 Q_3 , 
\end{equation}
or, in terms of the  original magic simplex parameters
\begin{equation} \label{twoparameters}
3 c[0, 0] + c[0, 1] + 2 c[0, 2] + c[1, 1] + 2 c[1, 2] + c[2, 1] + 
 2 c[2, 2]   >2. 
\end{equation}
\subsubsection{Choi test}
Also, Example 2 in \cite{bae2018entanglement} states that the ``Choi EW $W^{(+)}$ obtained from the Choi map in $d=3$ \ldots is given by
\begin{equation}
W^{(+)} = \frac{1}{6} \left( \sum_{i=0}^{2} [ 2| ii \rangle \langle ii |  + | i,i-1 \rangle \langle i, i-1 | ] - 3 \mathrm{P}_{+}  \right) \nonumber,
\end{equation}
where $\mathrm{P}_+ = |\phi^+\rangle \langle \phi^+|$ with the Bell state $|\phi^+\rangle = (|00\rangle + |11\rangle + |22\rangle) / \sqrt{3}$." It was noted there that this witness (zeros are denoted by dots)
\begin{equation} \label{ChoiWitness}
W^{(+)} = \frac{1}{6} \left(\mathcode`0=\cdot
\begin{array}{ *{3}{c} | *{3}{c} | *{3}{c} }
 1 & 0 & 0 & 0 & -1 & 0 & 0 & 0 & -1 \\
 0 & 0 & 0 & 0 & 0 & 0 & 0 & 0 & 0 \\
 0 & 0 & 1 & 0 & 0 & 0 & 0 & 0 & 0 \\\hline
 0 & 0 & 0 & 1 & 0 & 0 & 0 & 0 & 0 \\
 -1 & 0 & 0 & 0 & 1 & 0 & 0 & 0 & -1 \\
 0 & 0 & 0 & 0 & 0 & 0 & 0 & 0 & 0 \\\hline
 0 & 0 & 0 & 0 & 0 & 0 & 0 & 0 & 0 \\
 0 & 0 & 0 & 0 & 0 & 0 & 0 & 1 & 0 \\
 -1 & 0 & 0 & 0 & -1 & 0 & 0 & 0 & 1 \\
\end{array}
\right)   
\end{equation}
is applicable in the two-qutrit setting. (A related operator $W^{(-)}$ is presented in \cite{bae2018entanglement}, but it is decomposable and, thus, not capable of detecting bound entanglement. Further, in our particular analysis here, no entanglement at all was detected with its use.)
In the Hiesmayr-L{\"o}ffler $d=3$ two-qutrit 
density-matrix setting (\ref{densitymatrix}), the entanglement requirement that $\mbox{Tr} [W \rho_3]<0$ takes the simple form 
\begin{equation} \label{Choi}
2 Q_3+1-2 Q_1-3 Q_2 <0,    
\end{equation}
or, in terms of the  original magic simplex parameters,
\begin{equation}
1/6 (-c[0, 0] + c[0, 2] + 2 c[1, 0] + c[1, 2] + 2 c[2, 0] + c[2, 2])   <0.
\end{equation}
\subsection{Two-qutrit analyses ($d=3$)}
In Table~\ref{tab:Main}, we report certain computations  based on the preceding constraints. 
Hopefully, these provide insight into the underlying (Hilbert-Schmidt) geometry of the Hiesmayr-L{\"o}ffler $d=3$ magic simplex model.

\begin{figure}
    \centering
    \includegraphics{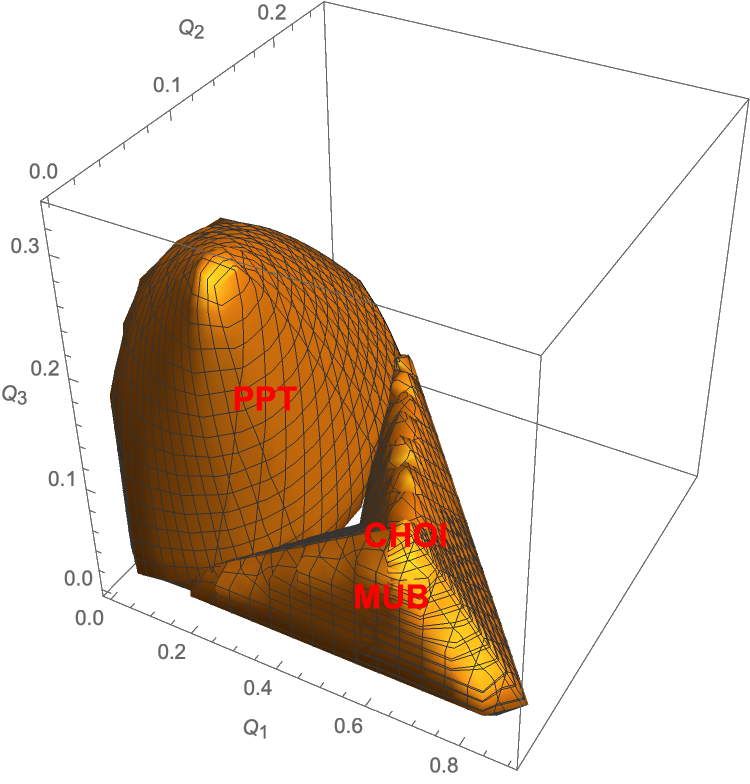}
    \caption{The largest object is the convex body of PPT states, while the more forward of the other two equal-sized (intersecting) objects is composed of those entangled states revealed by the MUB test, and the more backward-positioned one, composed of those entangled states revealed by the Choi test.}
    \label{fig:PPTMUBCHOI}
\end{figure}

\begin{table} 
\begin{center}
 \begin{tabular}{||c c c ||} 
 \hline
 Constraint Imposed & Probability  & Numerical Value  \\ 
 \hline
 \hline
 -------& 1 & 1. \\ 
 \hline
 PPT & $\frac{8 \pi }{27 \sqrt{3}}$ & 0.537422  \\
 \hline
 MUB & $\frac{1}{6}$ & 0.1666667  \\
 \hline
Choi & $\frac{1}{6}$ & 0.1666667  \\
 \hline
 $\text{PPT}\land \text{MUB}$ & $-\frac{4}{9}+\frac{4 \pi }{27 \sqrt{3}}+\frac{\log (3)}{6}$ & 0.00736862 \\
 \hline
 $\text{PPT}\land \text{Choi}$ & $-\frac{4}{9}+\frac{4 \pi }{27 \sqrt{3}}+\frac{\log (3)}{6}$ & 0.00736862 \\
 \hline
 $\text{MUB}\land \text{Choi}$ & $\frac{1}{9}$ & 0.11111 \\
 \hline
 $\text{MUB}\lor \text{Choi}$ & $\frac{2}{9}$ & 0.22222 \\
  \hline
   $\neg \text{MUB}\land \text{Choi}$ & $\frac{1}{18}$ & 0.05555 \\
    \hline
   $\text{MUB}\land  \neg \text{Choi}$ & $\frac{1}{18}$ & 0.05555 \\
 \hline
 $\text{PPT}\land \neg \text{MUB}$ & $\frac{1}{162} \left(72+8 \sqrt{3} \pi -27 \log (3)\right)$ & 0.5300534 \\
 \hline
 $\text{PPT}\land \neg \text{Choi}$ & $\frac{1}{162} \left(72+8 \sqrt{3} \pi -27 \log (3)\right)$ & 0.5300534 \\
 \hline
$\text{PPT}\land \text{MUB}\land \text{Choi}$ & 0 & 0 \\
\hline
$\text{PPT}\land(\text{MUB}\lor \text{Choi})$ & $-\frac{8}{9}+\frac{8 \pi }{27 \sqrt{3}}+\frac{\log (3)}{3}$ & 0.0147372 \\
\hline
$\neg \text{PPT}\land \text{MUB}$ & $\frac{1}{3}+\frac{22518 \sqrt{3}}{91}+\frac{3888 \sqrt{3}}{7 \pi }-\frac{10939 \pi }{27
   \sqrt{3}}-\frac{\log (3)}{8}$ & 0.1592980 \\
\hline
$\neg \text{PPT}\land \text{Choi}$ & $\frac{1}{3}+\frac{22518 \sqrt{3}}{91}+\frac{3888 \sqrt{3}}{7 \pi }-\frac{10939 \pi }{27
   \sqrt{3}}-\frac{\log (3)}{8}$ & 0.1592980 \\
  \hline
 $\neg \text{PPT}\land \neg \text{MUB}$ &$\frac{1}{162} \left(9 (7+\log (27))-8 \sqrt{3} \pi \right)$ & 0.303279920
 \\
 \hline
  \hline
 $\neg \text{PPT}\land \neg \text{Choi}$ &$\frac{1}{162} \left(9 (7+\log (27))-8 \sqrt{3} \pi \right)$ & 0.303279920
 \\
 \hline
 $\neg \text{PPT}\land \neg \text{MUB} \land \neg \text{Choi}$ & $\frac{1}{9} (3 \log (3)-1)$ & 0.255092985 \\
  \hline
 $ \text{PPT}\land \neg \text{MUB} \land \neg \text{Choi}$ & $\frac{1}{9} (8-3 \log (3))$ & 0.5226847927 \\
 \hline
 $\text{PPT}\lor (\text{MUB}\land \text{Choi})$ & $\frac{1}{81} \left(9+8 \sqrt{3} \pi \right)$ & 0.648533145 \\
 \hline
\end{tabular}
\caption{\label{tab:Main}Various Hilbert-Schmidt probabilities for the Hiesmayr-L{\"o}ffler $d=3$ two-qutrit model. Notationally, $\neg$ is the negation logic operator (NOT); $\land$ is the conjunction logic operator (AND); and $\lor$  is the disjunction logic operator (OR).}
\end{center}
\end{table}
To begin, we obtained findings (Fig.~\ref{fig:PPTMUBCHOI}, cf. \cite{kglr}) for the $d=3$ two-qutrit Hiesmayr-L{\"o}ffler model that 
the PPT-probability  is $\frac{8 \pi }{27 \sqrt{3}} \approx 0.537422$, and that the entanglement probabilities revealed by the MUB and Choi tests are both $\frac{1}{6} \approx 0.16667$. 

Further, the MUB and Choi bound-entangled probabilities were each determined to equal $-\frac{4}{9}+\frac{4 \pi }{27 \sqrt{3}}+\frac{\log (3)}{6} \approx  0.00736862$. 
So, the intersection of the PPT-set with either of the two sets  accounts for less than one-percent of the total probability. (This smallness appears to be very much in line with that exhibited in Figure 3 in \cite{hiesmayr2013complementarity}--also \cite[Figs. 2, 3]{baumgartner2006state}.)

If we enforce both the MUB and Choi tests for entanglement, but not the PPT constraint, we obtain a probability of 
$\frac{1}{9} \approx 0.11111$. This  doubles to $\frac{2}{9} \approx 0.22222$, if only one of the two tests needs to be met,  again without PPT necessarily holding. (``An entanglement witness is an observable detecting entanglement for a subset of states. We present a framework that makes an entanglement witness {\it twice} [emphasis added] as powerful due to the general existence of a second (lower) bound, in addition to the (upper) bound of the very definition" \cite{bae2018entanglement}.) Continuing, then, the probability for either of the tests to be met, but not the other, is $\frac{1}{18} \approx 0.05555$.  

If, on the other hand, the PPT-constraint is  satified, but both of the two entanglement tests are failed, the associated probability is  $\frac{1}{9} (8-3 \log (3)) \approx  0.5226847927$.

The intersection is void between those (bound-entangled) states satisfying both PPT and MUB criteria and those (bound-entangled) states satisfying both PPT and Choi criteria. (In their two-qutrit analysis, Gabuldin and Mandilara concluded that the bound-entangled states had ``negligible volume and that these form tiny `islands' sporadically distributed over the surface of the polytope of separable states" \cite{gabdulin2019investigating}. In a continuous variable study \cite{diguglielmo2011experimental}, ``the tiny regions in parameter space where bound entanglement does exist'' were noted'' )

\begin{figure}
    \centering
    \includegraphics{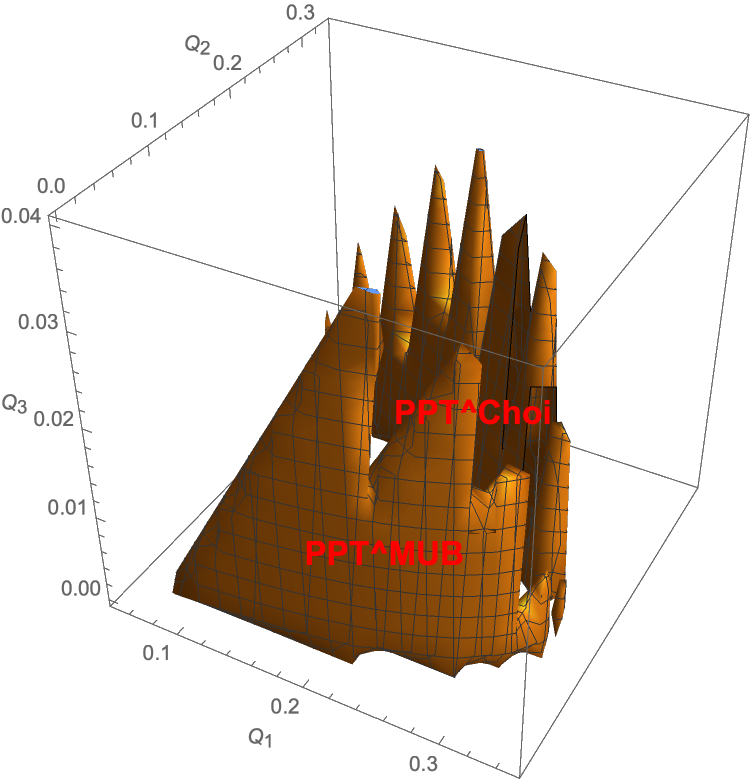}
    \caption{Jagged islands of bound-entangled states. The  $\text{PPT}\land \text{MUB}$ island is situated in front of the  $\text{PPT}\land \text{Choi}$ one. The former is confined to $0 <Q_2<\frac{1}{9}$ and the latter to $\frac{1}{9}< Q_2<\frac{1}{3}$.}
    \label{fig:Void}
\end{figure}

In Fig.~\ref{fig:Void} we show these non-intersecting ragged/craggy/jagged islands of bound-entangled states. (The notion of fractal-type behavior comes to mind.) The bound-entangled probabilities of  $-\frac{4}{9}+\frac{4 \pi }{27 \sqrt{3}}+\frac{\log (3)}{6} \approx 0.00736862$ are equal to 36 times the depicted volumes.

\begin{figure}
    \centering
    \includegraphics{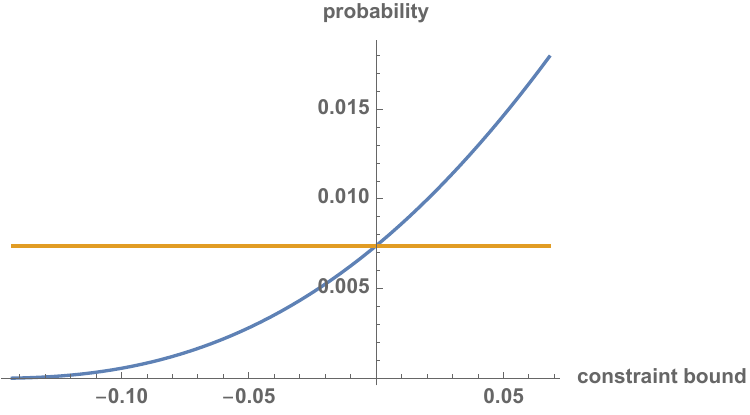}
    \caption{Probability as a function of the constraint bound $\mbox{Tr} [W \rho_3]$ imposed on the Choi witness. When the bound is zero--as in (\ref{Choi})--the indicated bound-entangled probability of 
    $-\frac{4}{9}+\frac{4 \pi }{27 \sqrt{3}}+\frac{\log (3)}{6} \approx  0.00736862$ is obtained. }
    \label{fig:VariableConstraint}
\end{figure}

\begin{figure}
    \centering
    \includegraphics{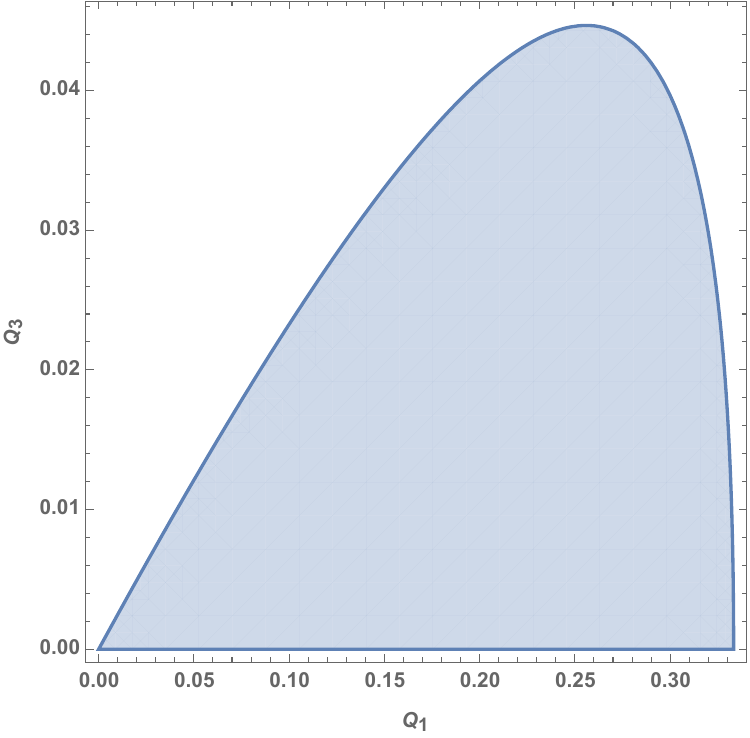}
    \caption{Points situated at the boundary between the separable and bound-entanglement Hiesmayr-L{\"o}ffler $d=3$ two-qutrit states, based on the MUB constraint (\ref{MUBconstraint}), by setting $Q_2=\frac{1}{3} (Q_1-4 Q_3)$}
    \label{fig:BoundaryPoints}
\end{figure}

In Fig.~\ref{fig:VariableConstraint} we show the effect upon the calculated probability of allowing the right-hand side (zero) of the constraint (\ref{Choi}) to instead vary.
 When the bound is zero--as in (\ref{Choi})--the indicated bound-entangled probability of 
    $-\frac{4}{9}+\frac{4 \pi }{27 \sqrt{3}}+\frac{\log (3)}{6} \approx  0.00736862$ is obtained. If the bound is further loosened to  as high as $\frac{43}{32} \approx 1.34375$, where the associated constraint loses all effect, the curve attains the PPT-probability of $\frac{8 \pi }{27 \sqrt{3}} \approx 0.537422$. At the other extreme, for a bound as strong as $-\frac{3}{16} \approx -0.1875$, the probability yielded is simply zero.
In Fig.~\ref{fig:BoundaryPoints} we show the points lying at the boundary between the separable and bound-entanglement states based on the MUB witness, obtained by setting $Q_2=\frac{1}{3} (Q_1-4 Q_3)$, so that the inequality constraint (\ref{MUBconstraint}) becomes an equality.
\subsubsection{Extension of Choi witness $W^{(+)}$ to $W^{(+)}(a)$} \label{ChoiExtension}
From the papers \cite{chruscinski2018entanglement,ha2011one}
we obtained a general (initially three-parameter ($[a,b,c]$) set of entanglement witnesses $W^{(+)}(a)$ (with the Choi witness already employed corresponding to $a=1$). It was generated by replacing (making use of cyclical permutations of $a,b,c$) the diagonal entries of the witness (\ref{ChoiWitness}) by $\{a,b,c,c,a,b,b,c,a\}$. The conditions imposed on the three parameters are \cite[eq. (1)]{ha2011one}
 \begin{equation}
 a+b+c=2, 0 \leq a \leq 1, b,c \geq 0 , b c =(1-a)^2 .
 \end{equation}
 
 To start our associated analyses, for $a=\frac{1}{4} \left(3-\sqrt{5}\right)$, we found an entanglement probability of $\frac{5}{132} \left(5+\sqrt{5}\right) \approx 0.274093$ and a bound-entangled probability of 0.00149772192.
 
Further, we note that $a=\frac{1}{3}$ is the particular case $i=4$ of \cite[eq. (1)]{ha2011one}
\begin{equation}
 a=   \frac{2}{3} \left(\cos \left(\frac{\pi  i}{3}\right)+1\right),  b= \frac{2}{3}
   \left(-\frac{1}{2} \sqrt{3} \sin \left(\frac{\pi  i}{3}\right)-\frac{1}{2} \cos
   \left(\frac{\pi  i}{3}\right)+1\right),
\end{equation}
and
\begin{displaymath}
 c=\frac{2}{3} \left(\frac{1}{2} \sqrt{3} \sin
   \left(\frac{\pi  i}{3}\right)-\frac{1}{2} \cos \left(\frac{\pi 
   i}{3}\right)+1\right).
\end{displaymath}
For $1 \le i \leq 5$ the  corresponding witness is optimal \cite{ha2011one}.

Now, for $a=\frac{1}{3}$,  we have an entanglement probability of $\frac{125}{486} =\frac{5^3}{2 \cdot 3^5} \approx 0.257202$, with
a bound-entangled probability of 
\begin{equation}
 \frac{-204+7 \log (7)+168 \sqrt{3} \cos ^{-1}\left(\frac{11}{14}\right)}{1134}   \approx 0.00325612294236.
\end{equation}

The case $a=1$ yields back our initial Choi witness $W^{(+)}$ analysis, for which a bound-entangled probability of $-\frac{4}{9}+\frac{4 \pi }{27 \sqrt{3}}+\frac{\log (3)}{6} \approx  0.00736862$ was obtained.

\begin{figure}
    \centering
    \includegraphics{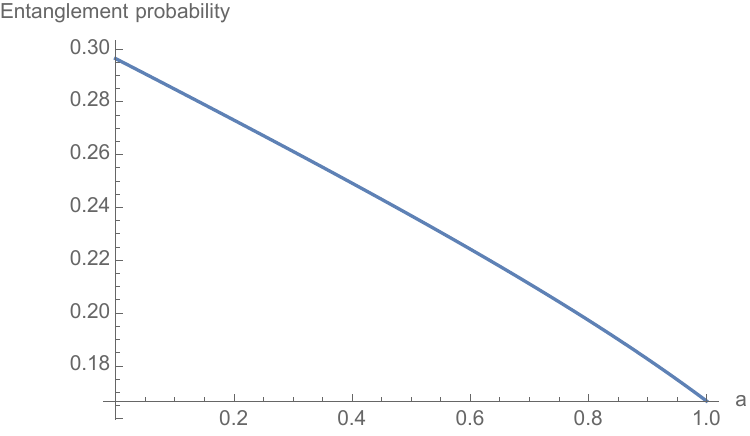}
    \caption{Probability of entanglement--$-\frac{(a-2)^3}{9 a^2-30 a+27}$--obtained by application of the family of generalized Choi witnesses $W^{(+)}(a)$ to the Hiesmayr-L{\"o}ffler $d=3$ two-qutrit model. The maximum of $\frac{8}{27} \approx 0.296296$ is attained at $a=0$.}
    \label{fig:EntangledProbability}
\end{figure}

\begin{figure}
    \centering
    \includegraphics{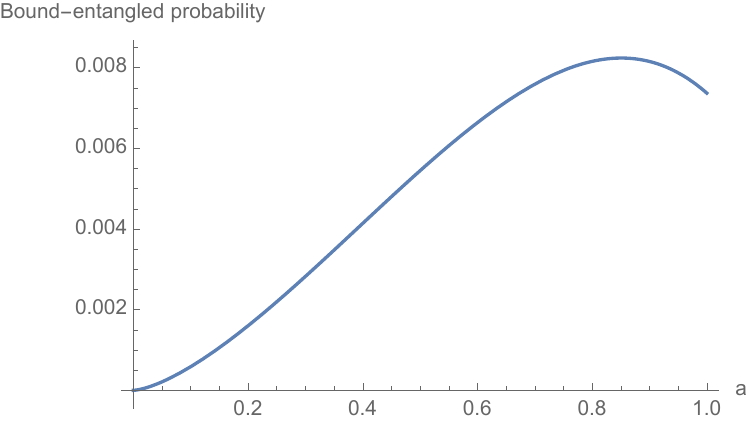}
    \caption{Probability of bound entanglement (\ref{Tessore}) obtained by application of the family of generalized Choi witnesses $W^{(+)}(a)$ to the Hiesmayr-L{\"o}ffler $d=3$ two-qutrit model. The peak  $\approx 0.0082381739$  is near  $a= 0.8509958326$. (These two values are quite complicated analytically.) At $a=\frac{1}{3}$, the probability is $\frac{-204+7 \log (7)+168 \sqrt{3} \cos ^{-1}\left(\frac{11}{14}\right)}{1134}  \approx 0.00325612294236$. At $a=1$, the probability is $-\frac{4}{9}+\frac{4 \pi }{27 \sqrt{3}}+\frac{\log (3)}{6} \approx 0.00736862$ (Table~\ref{tab:Main}).}
    \label{fig:BoundEntangledProbPlot}
\end{figure}

More, generally still, the entanglement probabilities for this class of witnesses $W^{(+)}(a)$, are given by (Fig.~\ref{fig:EntangledProbability})
\begin{equation} \label{EntanglementProbability}
 -\frac{(a-2)^3}{9 a^2-30 a+27}.   
\end{equation}
\paragraph{\underline{First one-parameter family of bound-entangled probabilities.}}
Further, the bound-entangled probability   obtained by application of the family of generalized Choi witnesses $W^{(+)}(a)$ to the Hiesmayr-L{\"o}ffler $d=3$ two-qutrit model is given by the expression (Fig.~\ref{fig:BoundEntangledProbPlot})
\begin{equation} \label{Tessore}
-\frac{A+B}{54 (4-3 a)^{3/2} (2 a-3)} ,   
\end{equation} 
where
\begin{displaymath}
A=8 \sqrt{12-9 a} \left(6 a^2-17 a+12\right) \cos ^{-1}\left(\frac{a (3 a-8)+6}{6-4
   a}\right)
\end{displaymath}
and
\begin{displaymath}
B=3 \sqrt{a} \left(2 \left(9 a^3-57 a^2+108 a-64\right)+3 (3-2 a) a \log (9-6 a)\right).
\end{displaymath}

In regard to this formula (\ref{Tessore}), C. Dunkl remarked: ``I have one intuitive observation: the mixture of special functions (including trig) in the answers appears to imply that the boundaries of the sets you are measuring are complicated and have pieces of various properties (e.g. flat, curved ...)". As an expansion upon this remark (cf. Fig.~\ref{fig:Void})--suggestive of the jagged-island phenomenon--let us note that the bound-entangled probability function 
(\ref{Tessore}) was obtained by the integration of the value 36, firstly  of $Q_2$ over the interval $[\frac{1}{6} \left(-Q_1-2 Q_3+1\right)-\frac{1}{6} \sqrt{-3 Q_1^2+12 Q_3 Q_1-2 Q_1-4
   Q_3+1},\frac{1}{6} \left(-Q_1-2 Q_3+1\right)-\frac{1}{6} \sqrt{-\frac{\left((a-2) \left(3
   Q_1-1\right)+6 a Q_3\right){}^2}{a (3 a-4)}}]$, secondly of $Q_3$ over the interval $[0,\frac{a \left(\sqrt{\frac{(3 a-4) \left(3 Q_1-1\right) \left(3 (3 a-4)
   Q_1+4\right)}{a^2}}+3\right)+(15-9 a) Q_1-5}{9 a}]$, and thirdly of $Q_1$ over $[\frac{-a^2+2 a-1}{2 a-3},\frac{1}{3}]$.

\begin{figure}
    \centering
    \includegraphics{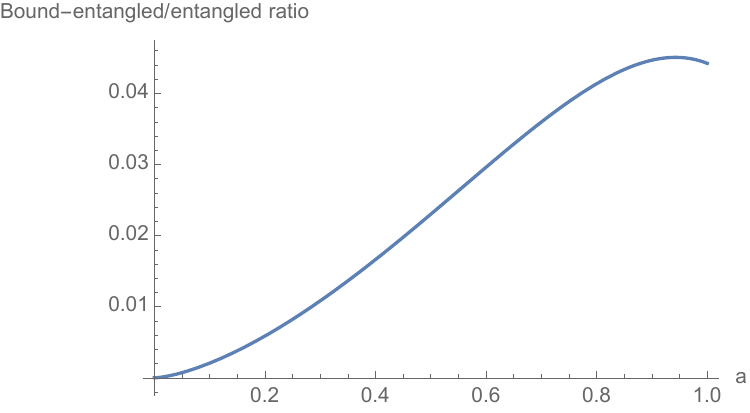}
    \caption{Ratio of bound-entangled probability to entanglement probability for the Hiesmayr-L{\"o}ffler $d=3$ two-qutrit model in terms of the generalized Choi entanglement witness $W^{(+)}(a)$. The peak  $\approx 0.0450501588$ is near  $a= 0.94280530186$.}
    \label{fig:BoundEntangledProbRatio}
\end{figure}

In Fig.~\ref{fig:BoundEntangledProbRatio}, we display the ratio of the bound-entangled probability (\ref{Tessore}) to the entanglement probability (\ref{EntanglementProbability}) .
\subsubsection{Entanglement witnesses from mutually unbiased bases}
Moving on to further forms of witnesses, we observed that for the entanglement witness given in \cite[eq. (32)]{chruscinski2018entanglement},
\begin{equation}
\frac{1}{3} \left(\mathcode`0=\cdot
\begin{array}{ *{3}{c} | *{3}{c} | *{3}{c} }
 4 & 0 & 0 & 0 & -1 & 0 & 0 & 0 & -1 \\
 0 & 1 & 0 & 0 & 0 & 2 & 2 & 0 & 0 \\
 0 & 0 & 1 & 2 & 0 & 0 & 0 & 2 & 0 \\\hline
 0 & 0 & 2 & 1 & 0 & 0 & 0 & 2 & 0 \\
 -1 & 0 & 0 & 0 & 4 & 0 & 0 & 0 & -1 \\
 0 & 2 & 0 & 0 & 0 & 1 & 2 & 0 & 0 \\\hline
 0 & 2 & 0 & 0 & 0 & 2 & 1 & 0 & 0 \\
 0 & 0 & 2 & 2 & 0 & 0 & 0 & 1 & 0 \\
 -1 & 0 & 0 & 0 & -1 & 0 & 0 & 0 & 4 \\
\end{array}
\right)  ,  
\end{equation}
there was no entanglement at all detected for the $d=3$ Hiesmayr-L{\"o}ffler model under study here. 

This last witness is the particular case $\phi_1=\phi_2= \pi, \phi_3=\phi_4 =0$ of a family parameterized by an $L$-dimensional torus ($L=2,3,4$, being the number of MUBs' used in the construction).

As an exercise, we randomly generated values of $\phi_1, \phi_2, \phi_3$ and $\phi_4$ $\in [0,2 \pi]$--and evaluated their (bound and non-bound/free) entanglement probabilities with respect to the $d=3$ Hiesmayr-L{\"o}ffler model. Neither of the probabilities generated exceeded the maximum ones indicated ($\frac{8}{27} \approx 0.296296$ and  $\approx 0.0082381739$ in Figs.~\ref{fig:EntangledProbability} and \ref{fig:BoundEntangledProbPlot}).
\subsubsection{Jafarizadeh-Behzadi-Akbari witnesses} \label{IranianTwoQutrits}
In \cite{jafarizadeh2009entanglement}, M. A. Jafarizadeh, N. Behzadi, and Y. Akbari (JBA), constructed ``[o]n the basis of linear programming, new sets of entanglement witnesses (EWs) for $3 \otimes 3$
and $4 \otimes 4$ systems". In the $3 \otimes 3$ case, their two witnesses $\mathcal{W}_{\alpha}$ and $\mathcal{W'}_{\alpha}$ were defined over $\alpha \in [\frac{1}{3},\frac{2}{3}]$ \cite[eqs. (17), (18)]{jafarizadeh2009entanglement}.

The associated entanglement constraints for the $d=3$ Hiesmayr-L{\"o}ffler model took the forms
\begin{equation} \label{Iranian1}
\alpha  \left(-3 \alpha  Q_1+(9 \alpha -3) Q_2+6 \alpha  Q_3+Q_1\right)<0,
\end{equation}
and 
\begin{equation}\label{Iranian2}
\alpha  (3 \alpha -1) \left(2 Q_1+3 Q_2-1\right)>2 \alpha  Q_3,
\end{equation}
respectively.

\begin{figure}
    \centering
    \includegraphics{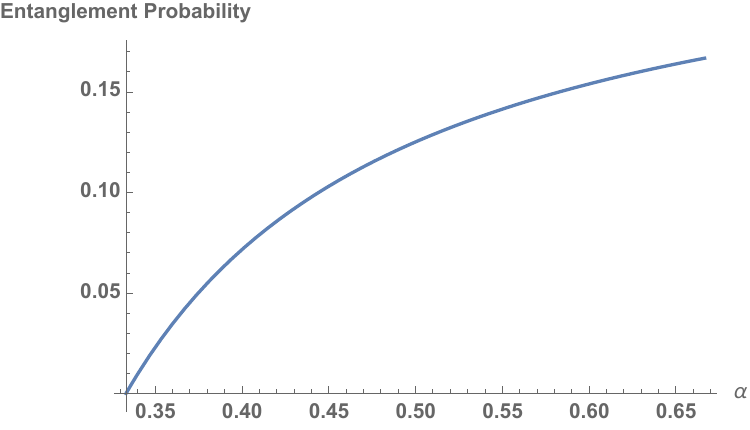}
    \caption{Identical entanglement probabilities, $\frac{1-3 \alpha }{2-12 \alpha }$, over $\alpha \in [\frac{1}{3},\frac{2}{3}]$, for the $d=3$ Hiesmayr-L{\"o}ffler model, based on either JBA witness (\ref{Iranian1}) or  (\ref{Iranian2}).}
    \label{fig:IranianEntanglement}
\end{figure}

The entanglement probabilities for the two witnesses were identical (Fig.~\ref{fig:IranianEntanglement}), being given by
\begin{equation}
\frac{1-3 \alpha }{2-12 \alpha }.
\end{equation}
\paragraph{\underline{Second one-parameter family of bound-entangled probabilities.}}

\begin{figure}
    \centering
    \includegraphics{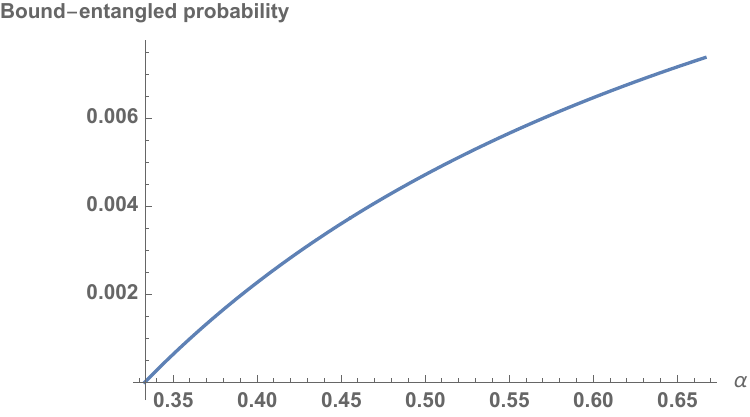}
    \caption{Bound-entangled probability function (\ref{JBAbound3D}) for the $d=3$ Hiesmayr-L{\"o}ffler model based on either  JBA witness $\mathcal{W}_{\alpha}$ or  $\mathcal{W'}_{\alpha}$. At $\alpha =\frac{1}{2}$, the probability is $\frac{1}{324} \left(-342+16 \sqrt{3} \pi +375 \sqrt{6}
   \text{csch}^{-1}\left(\frac{5}{\sqrt{2}}\right)\right) \approx 0.00470668$, while at $\alpha =\frac{2}{3}$, it is $\frac{1}{162} \left(8 \sqrt{3} \pi +9 (\log (27)-8)\right) \approx 0.00736862$.}
    \label{fig:IranianBound}
\end{figure}

\begin{figure}
    \centering
    \includegraphics{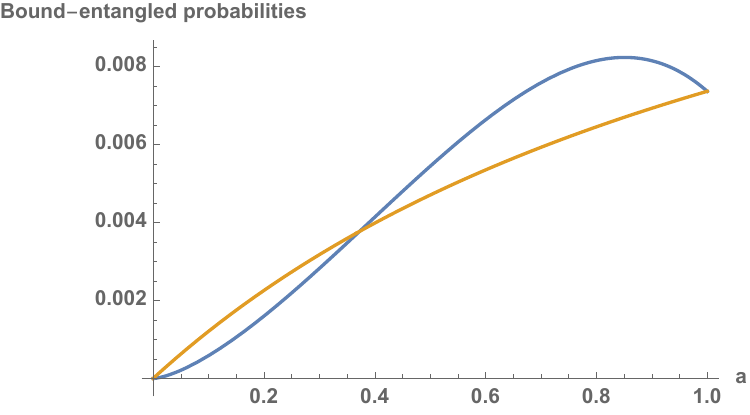}
    \caption{Rescaled ($\alpha \to \frac{a+1}{3}$) JBA-based Fig.~\ref{fig:IranianBound} plotted along with generalized-Choi-based Fig.~\ref{fig:BoundEntangledProbPlot}. The curves intersect at $a \approx 0.372519577$ with a common bound-entangled probability $\approx 0.00377346692$.}
    \label{fig:IntersectionBound}
\end{figure}

The bound-entangled probability  (Fig.~\ref{fig:IranianBound}) based on 
either of these two JBA witnesses is given
by the product of 
\begin{equation} \label{JBAbound3D}
  \frac{1}{162 (6 (1-3 \alpha ) \alpha +1)^2}  
\end{equation}
and
\begin{displaymath}
8 \sqrt{3} \pi  (6 (1-3 \alpha ) \alpha +1)^2-18 (3 \alpha -1) (15 \alpha +2) (6 \alpha 
   (3 \alpha -1)-1)
\end{displaymath}
\begin{displaymath}
-6 \sqrt{18 (1-3 \alpha ) \alpha +3} (3 \alpha +1)^3 \csc
   ^{-1}\left(\frac{3 \alpha +1}{\sqrt{6 (1-3 \alpha ) \alpha +1}}\right).
\end{displaymath}
In Fig.~\ref{fig:IntersectionBound}, we jointly plot a rescaled version of this bound-entangled probability figure based on the JBA witnesses along with that earlier-derived one (Fig.~\ref{fig:BoundEntangledProbPlot}) using the generalized Choi witnesses.

\begin{figure}
    \centering
    \includegraphics{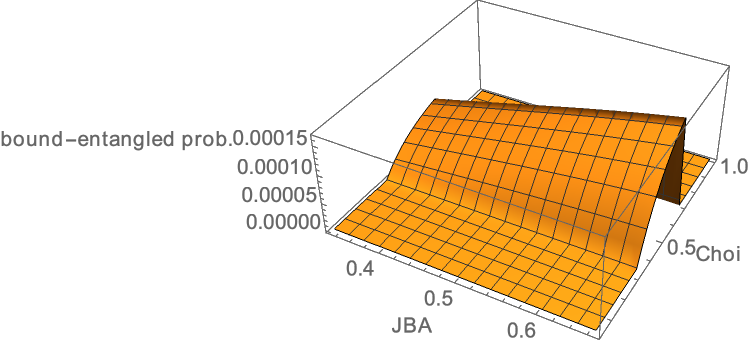}
    \caption{Probabilities of the {\it intersections}  of the bound-entangled regions revealed through use of the one-parameter ($ \alpha \in[\frac{1}{3}, \frac{2}{3}], a\in [0,1]$) families of generalized JBA and Choi witnesses ($d=3$)}
    \label{fig:ChoiJBAIntersection}
\end{figure}

\begin{figure}
    \centering
    \includegraphics{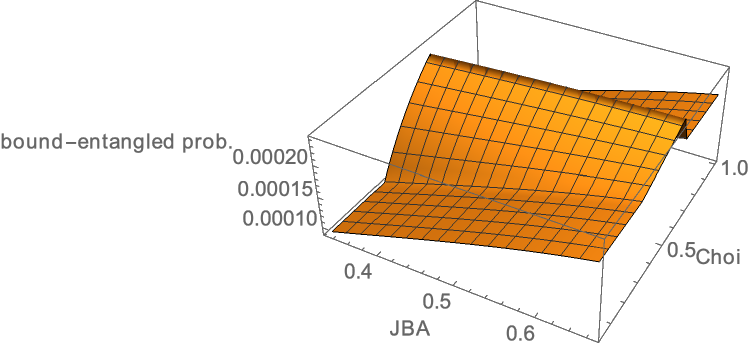}
    \caption{Probabilities of the {\it unions} of the bound-entangled regions revealed through use of the one-parameter ($\alpha \in [\frac{1}{3},\frac{2}{3}], a \in [0,1]$) families of generalized JBA and Choi witnesses ($d=3$)}
    \label{fig:ChoiJBAUnion}
\end{figure}

In Figs.~\ref{fig:ChoiJBAIntersection} and \ref{fig:ChoiJBAUnion}, we show the probabilities of the intersections and unions of the bound-entangled regions revealed through use of the one-parameter families of generalized JBA and Choi witnesses.

\begin{figure}
     \centering
     \includegraphics{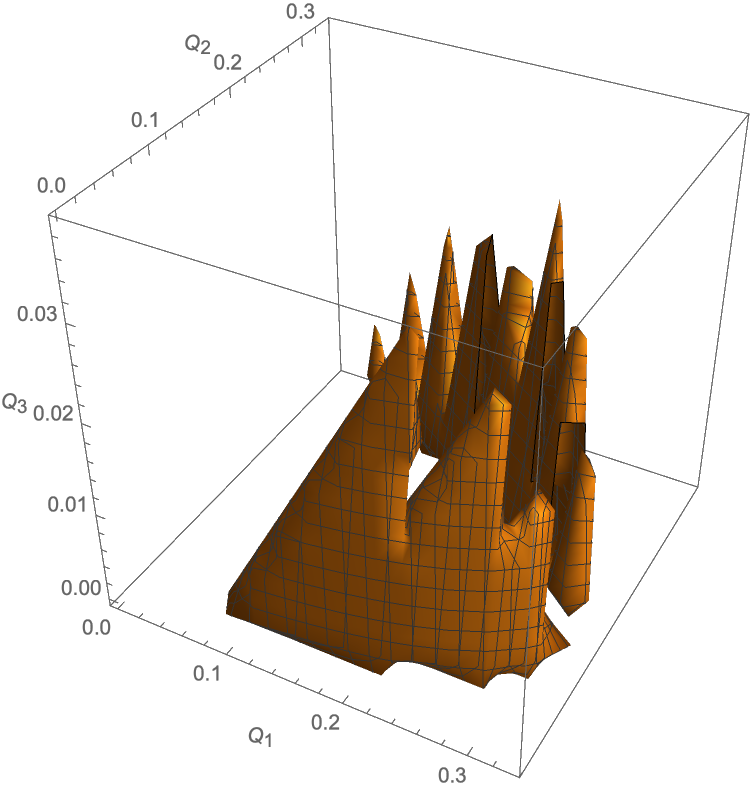}
     \caption{The anterior "jagged island" represents the  bound entanglement based on the Jafarizadeh-Behzadi-Akbari witness $\mathcal{W}_{\alpha}$ and the posterior one that based on $\mathcal{W'}_{\alpha}$ for the $d=3$ (two-qutrit) Hiesmayr-L{\"o}ffler magic simplex. The parameter $\alpha$ has been set to $\frac{3}{5}$.}
     \label{fig:IranianBoundEntanglementd3}
 \end{figure}

In Fig.~\ref{fig:IranianBoundEntanglementd3} we show a pair of jagged islands for $\alpha = \frac{3}{5}$.

It was commented upon in \cite{jafarizadeh2009entanglement} that the expectation value of $\mathcal{W}_{\alpha}$ ``with respect to the all separable states is positive hence it can be an EW for $\alpha \in [\frac{1}{3},\frac{2}{3}]$". The formula (\ref{JBAbound3D}) was derived assuming  these restrictions--but we have observed that if we take the limit $\alpha \rightarrow \infty$, it attains the value
\begin{equation}
 \frac{1}{324} \left(-90+16 \sqrt{3} \pi +3 \sqrt{6} \sinh
   ^{-1}\left(\sqrt{2}\right)\right) \approx    0.0169299.
\end{equation}
Possibly this is not an achievable (or near-achievable) bound-entanglement probability measurement. For $\alpha =\frac{2}{3}$, the formula  (\ref{JBAbound3D}) has the considerably lesser value $\frac{1}{162} \left(8 \sqrt{3} \pi +9 (\log (27)-8)\right) \approx 0.00736862$.
\subsection{Two-ququart analyses ($d=4$)} \label{Twoququarts} 
Similarly to our $d=3$ analyses (sec.~\ref{Transformation}), in the $d=4$ (two-ququart) framework, to transform between  the sixteen nonnegative parameters  ($c[k,l] \geq 0$, with $k,l=0,\ldots3$), summing to 1, for the full/original magic simplex \cite[sec. 4]{baumgartner2008geometry} and the four ($q_1,q_2, q_3, q_4$) of the Hiesmayr-L{\"o}fller subset, we must take
\begin{equation} \label{d=4Equality}
 c[0,0]=\frac{1}{880} \left(75 q_1-11 q_2-55 q_3-55 q_4+55\right) \equiv Q_1,
\end{equation} 
and
\begin{equation}
 c[k,l]=\frac{1}{880} \left(-5 q_1-11 q_2+165 q_3-55 q_4+55\right) \equiv Q_2
\end{equation} 
 for $\{k,l\}={\{0,1\},\{1,1\},\{2,1\}},\{3,1\}$, and
\begin{equation}
 c[k,l]=\frac{1}{880} \left(-5 q_1-11 q_2-55 q_3+165 q_4+55\right)  \equiv Q_3,
\end{equation}
 for $\{k,l\}={\{0,2\},\{1,2\},\{2,2\},\{3,2\}}$, and
\begin{equation}
 c[k,l]=\frac{-15 q_1+143 q_2-165 q_3-165 q_4+165}{2640} \equiv Q_4
\end{equation}
for $\{k,l\}={\{1,0\},\{2,0\},\{3,0\}}$.
 
If we employ $Q_1, Q_2$  $Q_3, Q_4$ as our four principal variables, rather than $q_1, q_2, q_3, q_4$ in the Hiesmayr-L{\"o}ffler parameterizaion (\ref{densitymatrix}), 
using the linear transformations,
\begin{equation}
q_1=  \frac{11}{4} \left(5 Q_1+4 Q_2+4 Q_3+3 Q_4-1\right), q_2= \frac{15}{4} \left(Q_1+4 Q_2+4 Q_3+7 Q_4-1\right),
\end{equation}
and 
\begin{displaymath}
q_3=Q_1+8 Q_2+4 Q_3+3 Q_4-1,   q_4=Q_1+4 Q_2+8 Q_3+3 Q_4-1,
\end{displaymath}
our ensuing analyses simplify greatly. The requirement that $\rho_4$ is a nonnegative definite density matrix--or, equivalently, that its sixteen leading nested minors are nonnegative--takes the form 
\begin{equation} \label{d=4Basic}
Q_1>0\land Q_4>0\land Q_2>0\land Q_3>0\land Q_1+4 \left(Q_2+Q_3\right)+3 Q_4<1.
\end{equation}
The constraint that the partial transpose of $\rho_4$ is nonnegative definite is 
\begin{equation} \label{d=4PPT}
Q_3>0\land Q_1+3 Q_4>0\land Q_1+4 \left(Q_2+Q_3\right)+3 Q_4<1\land Q_1^2+4 Q_2
   Q_1+Q_4^2  
\end{equation}
\begin{displaymath}
+16 Q_2 \left(Q_2+Q_3\right)+12 Q_2 Q_4<4 Q_2+2 Q_1 Q_4\land
   \left(Q_1-Q_4\right){}^2<16 Q_3^2 . 
\end{displaymath}
With these formulas, we are able to establish that the corresponding PPT-probability is $\frac{1}{2}+\frac{\log \left(2-\sqrt{3}\right)}{8 \sqrt{3}} \approx 0.404957$ (again, quite elegant, but seemingly of a different analytic form than the $d=3$ counterpart of 
$\frac{8 \pi}{27 \sqrt{3}}$). 

We were not able originally to compute bound-entangled probabilities in this  two-ququart framework, not being successful in attempting to extend the Hiesmayr-L{\"o}ffler and Choi witnesses to that setting. (``The general case (even for $d=4$) is much more involved and the general structure of circulant entanglement witnesses is not known" \cite{chruscinski2011geometry}.)
\subsubsection{ Chru{\'s}ci{\'n}ski witnesses}
However, Dariuz Chru{\'s}ci{\'n}ski subsequently  provided the particular entanglement witness.
\begin{equation} \label{DariuzMatrix}
W_{2ququarts}=
\left(\mathcode`0=\cdot
\begin{array}{ *{4}{c} | *{4}{c} | *{4}{c} | *{4}{c}}
 1 & 0 & 0 & 0 & 0 & -1 & 0 & 0 & 0 & 0 & -1 & 0 & 0 & 0 & 0 & -1 \\
 0 & 1 & 0 & 0 & 0 & 0 & 0 & 0 & 0 & 0 & 0 & 0 & 0 & 0 & 0 & 0 \\
 0 & 0 & 1 & 0 & 0 & 0 & 0 & 0 & 0 & 0 & 0 & 0 & 0 & 0 & 0 & 0 \\
 0 & 0 & 0 & 0 & 0 & 0 & 0 & 0 & 0 & 0 & 0 & 0 & 0 & 0 & 0 & 0 \\\hline
 0 & 0 & 0 & 0 & 0 & 0 & 0 & 0 & 0 & 0 & 0 & 0 & 0 & 0 & 0 & 0 \\
 -1 & 0 & 0 & 0 & 0 & 1 & 0 & 0 & 0 & 0 & -1 & 0 & 0 & 0 & 0 & -1 \\
 0 & 0 & 0 & 0 & 0 & 0 & 1 & 0 & 0 & 0 & 0 & 0 & 0 & 0 & 0 & 0 \\
 0 & 0 & 0 & 0 & 0 & 0 & 0 & 1 & 0 & 0 & 0 & 0 & 0 & 0 & 0 & 0 \\\hline
 0 & 0 & 0 & 0 & 0 & 0 & 0 & 0 & 1 & 0 & 0 & 0 & 0 & 0 & 0 & 0 \\
 0 & 0 & 0 & 0 & 0 & 0 & 0 & 0 & 0 & 0 & 0 & 0 & 0 & 0 & 0 & 0 \\
 -1 & 0 & 0 & 0 & 0 & -1 & 0 & 0 & 0 & 0 & 1 & 0 & 0 & 0 & 0 & -1 \\
 0 & 0 & 0 & 0 & 0 & 0 & 0 & 0 & 0 & 0 & 0 & 1 & 0 & 0 & 0 & 0 \\\hline
 0 & 0 & 0 & 0 & 0 & 0 & 0 & 0 & 0 & 0 & 0 & 0 & 1 & 0 & 0 & 0 \\
 0 & 0 & 0 & 0 & 0 & 0 & 0 & 0 & 0 & 0 & 0 & 0 & 0 & 1 & 0 & 0 \\
 0 & 0 & 0 & 0 & 0 & 0 & 0 & 0 & 0 & 0 & 0 & 0 & 0 & 0 & 0 & 0 \\
 -1 & 0 & 0 & 0 & 0 & -1 & 0 & 0 & 0 & 0 & -1 & 0 & 0 & 0 & 0 & 1 \\
\end{array}
\right)
\end{equation}
The constraint required for bound entanglement that $\mbox{Tr} [W_{2ququarts} \rho_4]<0$, then, takes the form
\begin{equation} \label{d=4BE}
2 \left(Q_2+Q_3\right)+3 Q_4<Q_1.  
\end{equation}
The Hilbert-Schmidt entanglement probability that the Hiesmayr-L{\o}ffler density matrix $\rho_4$--given by (\ref{densitymatrix})--satisfies this constraint is simply $\frac{2}{9}$.

Joining the constraints (\ref{d=4Basic}), (\ref{d=4PPT}) and (\ref{d=4BE}), we attempted the corresponding
exact four-dimensional integration for the bound-entangled probability. Mathematica was able to reduce it to a clearly challenging one-dimensional integration.
However, we were apparently able to obviate this formidable task by doing a numerical integration using a WorkingPrecision$\rightarrow 24$ option. Inputting
the result obtained to the WolframAlpha website, an exact value of 
\begin{equation} \label{firstDariuz}
 \frac{8 \log (2)}{27}-\frac{59}{288} \approx 0.00051583,   
\end{equation}
was suggested, which matched the numerical output to considerably more than twenty decimal places. (As a matter of, at least, initial curiosity, the WolframAlpha site also suggested--to equally high-precision--an exact value of $\frac{1}{864} (-64 b_{4}(2)-177)$, where ``$b_4(2)$ is the Madelung constant $b_4(2)$". Equating the two exact suggested values implies that $b_4(2) =-4 \log (2)$. We could not immediately confirm this identity, but the mathworld.wolfram.com Madelung Constants page informs us that, seemingly relatedly, $b_2(2)=- \pi \log (2)$.) At a later point, though, N. Tessore was able to fully confirm the validity of 
(\ref{firstDariuz}) through strictly symbolic integration
.

D. Chru{\'s}ci{\'n}ski also indicated that a modification of $W_{2ququarts}$ could be achieved by replacing the sixteen diagonal entries of (\ref{DariuzMatrix}) with the sequence  $\{2,1,0,0,0,2,1,0,0,0,2,1,1,0,0,2\}$. The associated entanglement constraint is, then,  
\begin{equation} \label{d=4BE2}
 4 Q_2 + 9 Q_4 < Q_1 .  
\end{equation} 
Doing so, leads to a reduced (from $\frac{2}{9}$) entanglement (without PPT required) probability of $\frac{1}{8}$, but a substantially increased bound-entangled probability of  \cite{VasilyMitch}
\begin{equation}  \label{secondDariuz}
    \frac{24 \text{csch}^{-1}\left(\frac{8}{\sqrt{17}}\right)}{17 \sqrt{17}}-\frac{91}{544} \approx 0.002187.
\end{equation}
This  is 4.24019 times greater than the bound-entangled probability, $\frac{8 \log (2)}{27}-\frac{59}{288}$, obtained with the initially used two-ququart witness (\ref{DariuzMatrix}). (For several efforts to graphically represent the results of the immediately preceding set of two-ququart analyses, along the lines of Fig.~\ref{fig:PPTMUBCHOI}, see \cite{kglr}--in particular, the animation there--obtained by variation of the $Q$'s--of MichaelE2.) 

The joint imposition of the two last constraints (\ref{d=4BE}) and (\ref{d=4BE2}) gives a (lesser) bound-entanglement probability of 
\begin{equation}
 -\frac{571}{2448}+\frac{5 \log (2)}{54}+\frac{24
   \text{csch}^{-1}\left(\frac{8}{\sqrt{17}}\right)}{17 \sqrt{17}} \approx    0.000395295.
\end{equation}

It would be of interest to embed this last pair of two-ququart entanglement witnesses provided by Chru{\'s}ci{\'n}ski into a parameterized family, similarly to the case of the two-qutrit entanglement witness $W^{(+)}(a)$ analyzed in sec.~\ref{ChoiExtension}. But to do so, however, seemed quite involved \cite[sec. 3]{chruscinski2012geometry}.

\begin{figure}
     \centering
     \includegraphics{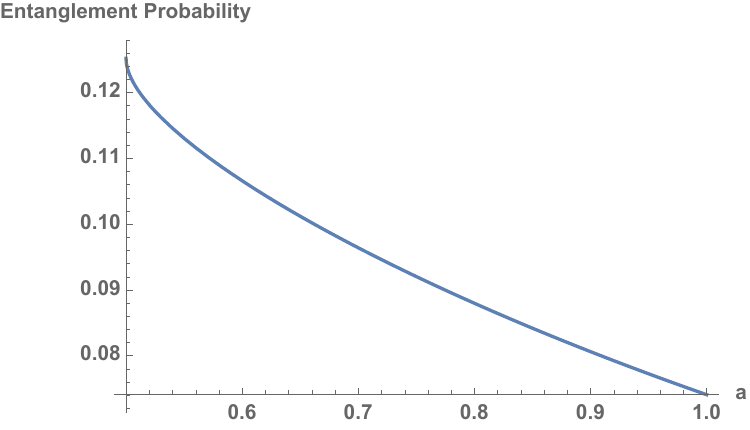}
     \caption{Entanglement probability--$\frac{(a-3)^4}{2 (a-7) (2 a-5) \left(-2 a+\sqrt{-4 (a-2) a-3}+7\right)}$ for class I of $W[a,b,c,d]$. At $a=\frac{1}{2}$, the probability is $\frac{625}{4992} \approx 0.1252$, and at $a=1$, it is $\frac{2}{27} \approx  0.0740741$.}
     \label{fig:Class1}
 \end{figure}

\begin{figure}
     \centering
     \includegraphics{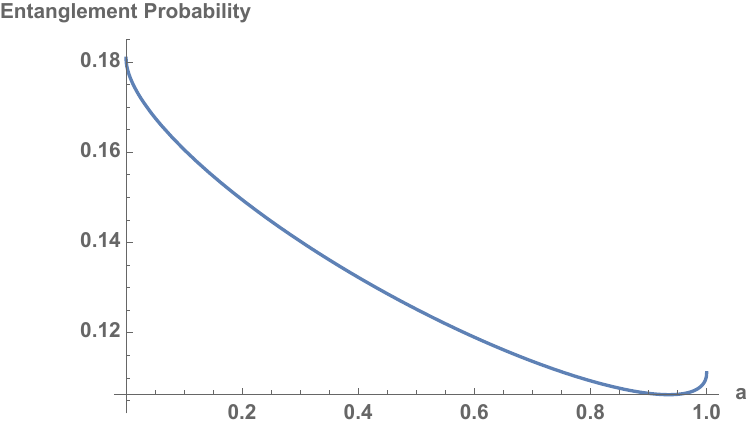}
     \caption{Entanglement probability--$ -\frac{(a-3)^4 \left(a+\sqrt{(1-a) a}-4\right)}{8 (a-7) (a-2) (a (2 a-9)+16)}$--for class II of $W[a,b,c,d]$. At $a=0$, the probability is $\frac{81}{448} \approx 0.180804$, and at $a=1$, it is $\frac{1}{9} \approx 0.11111$. The curve reaches a minimum of $\approx 0.1062629$ at $a \approx 0.9347153$.}
     \label{fig:Class2}
 \end{figure}

But, in \cite{chruscinski2012indecomposable}, we noted the presentation of a set of (nd-)optimal entanglement witnesses for $a, b, c, d \geq 0, a+b+c+d=3$ (cf. \cite[Example 5.2]{chruscinski2014class}),
\begin{equation} \label{W[a,b,c,d]}
 W[a,b,c,d]= \left(\mathcode`0=\cdot
\begin{array}{ *{4}{c} | *{4}{c} | *{4}{c} | *{4}{c}}
 a & 0 & 0 & 0 & 0 & -1 & 0 & 0 & 0 & 0 & -1 & 0 & 0 & 0 & 0 & -1 \\
 0 & b & 0 & 0 & 0 & 0 & 0 & 0 & 0 & 0 & 0 & 0 & 0 & 0 & 0 & 0 \\
 0 & 0 & c & 0 & 0 & 0 & 0 & 0 & 0 & 0 & 0 & 0 & 0 & 0 & 0 & 0 \\
 0 & 0 & 0 & d & 0 & 0 & 0 & 0 & 0 & 0 & 0 & 0 & 0 & 0 & 0 & 0 \\\hline
 0 & 0 & 0 & 0 & d & 0 & 0 & 0 & 0 & 0 & 0 & 0 & 0 & 0 & 0 & 0 \\
 -1 & 0 & 0 & 0 & 0 & a & 0 & 0 & 0 & 0 & -1 & 0 & 0 & 0 & 0 & -1 \\
 0 & 0 & 0 & 0 & 0 & 0 & b & 0 & 0 & 0 & 0 & 0 & 0 & 0 & 0 & 0 \\
 0 & 0 & 0 & 0 & 0 & 0 & 0 & c & 0 & 0 & 0 & 0 & 0 & 0 & 0 & 0 \\\hline
 0 & 0 & 0 & 0 & 0 & 0 & 0 & 0 & c & 0 & 0 & 0 & 0 & 0 & 0 & 0 \\
 0 & 0 & 0 & 0 & 0 & 0 & 0 & 0 & 0 & d & 0 & 0 & 0 & 0 & 0 & 0 \\
 -1 & 0 & 0 & 0 & 0 & -1 & 0 & 0 & 0 & 0 & a & 0 & 0 & 0 & 0 & -1 \\
 0 & 0 & 0 & 0 & 0 & 0 & 0 & 0 & 0 & 0 & 0 & b & 0 & 0 & 0 & 0 \\\hline
 0 & 0 & 0 & 0 & 0 & 0 & 0 & 0 & 0 & 0 & 0 & 0 & b & 0 & 0 & 0 \\
 0 & 0 & 0 & 0 & 0 & 0 & 0 & 0 & 0 & 0 & 0 & 0 & 0 & c & 0 & 0 \\
 0 & 0 & 0 & 0 & 0 & 0 & 0 & 0 & 0 & 0 & 0 & 0 & 0 & 0 & d & 0 \\
 -1 & 0 & 0 & 0 & 0 & -1 & 0 & 0 & 0 & 0 & -1 & 0 & 0 & 0 & 0 & a \\
\end{array}
\right)  
\end{equation}
Two classes of parameter constraints were considered,
\begin{equation}
   a+c =2, b+d=1, b d =(1-a)^2 
\end{equation}
and
\begin{equation}
 a+c=1, b+d=2, a c =(1-b)^2   .
\end{equation}
For neither of these classes, did we detect any bound entanglement with respect to the $d=4$ Hiesmayr-L{\"o}ffler system.
The entanglement probability (cf.(\ref{EntanglementProbability})) in class I does take the form (Fig.~\ref{fig:Class1}),
\begin{equation}
 \frac{(a-3)^4}{2 (a-7) (2 a-5) \left(-2 a+\sqrt{-4 (a-2) a-3}+7\right)}   
\end{equation}
and in  class II (Fig.~\ref{fig:Class2}),
\begin{equation}
 -\frac{(a-3)^4 \left(a+\sqrt{(1-a) a}-4\right)}{8 (a-7) (a-2) (a (2 a-9)+16)}.   
 \end{equation}
 These two functions are both equal to $\frac{625}{4992} \approx 0.1252$ at $a=\frac{1}{2}$. The intersection of the two entanglement tests at that 
 point, then, yields a probability of 
$ \frac{512}{4125} \approx 0.124121212$, while their union gives $\frac{512}{3135}  \approx 0.163317384$.

At $a=1$, the intersection of the two tests yields a probability of 
$ \frac{2}{16335} \approx 0.000122436$, while their union gives $\frac{32}{165}  \approx 0.193939$. At that same point, the case I entanglement probability  is  $\frac{2}{27} \approx  0.0740741$, and the case II entanglement probability is 
$\frac{1}{9} \approx 0.11111$.

\begin{figure}
    \centering
    \includegraphics{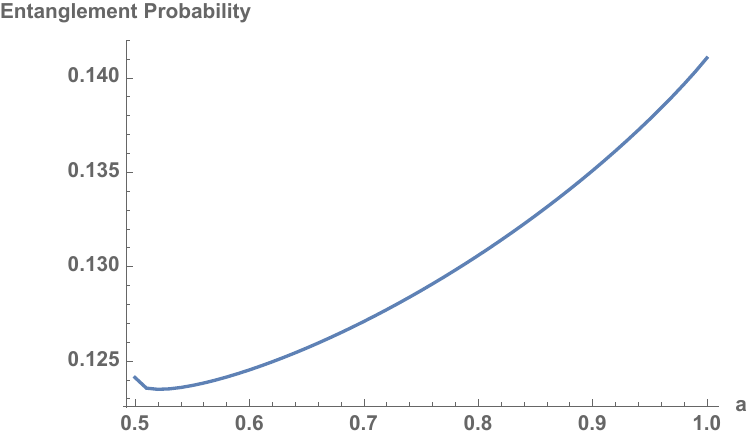}
    \caption{Entanglement probability for intersection of classes I and II for entanglement witness $W[a,b,c,d]$ (\ref{W[a,b,c,d]})}
    \label{fig:IntersectionPlot}
\end{figure}

\begin{figure}
    \centering
    \includegraphics{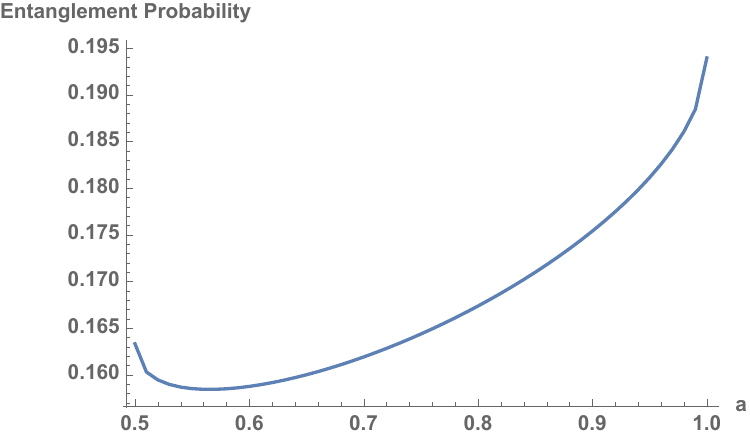}
    \caption{Entanglement probability for union of classes I and II for entanglement witness $W[a,b,c,d]$ (\ref{W[a,b,c,d]})}
    \label{fig:UnionPlot}
\end{figure}

\begin{figure}
    \centering
    \includegraphics{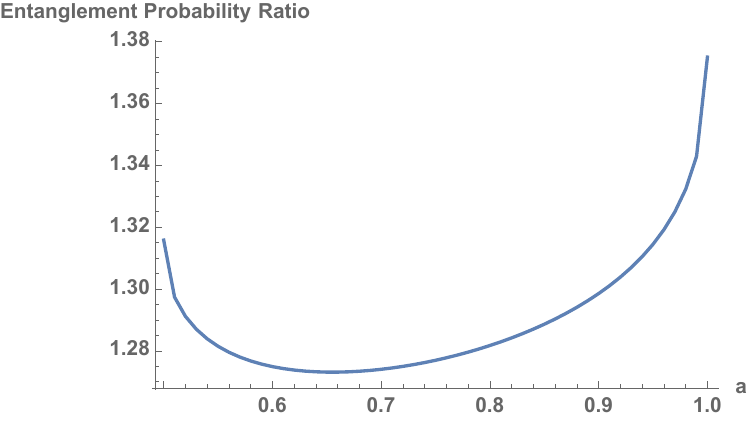}
    \caption{Ratio of union to intersection of classes I and II for entanglement witness $W[a,b,c,d]$ (\ref{W[a,b,c,d]})}
    \label{fig:UnionIntersectionRatio}
\end{figure}

In Figs.~\ref{fig:IntersectionPlot} and \ref{fig:UnionPlot}, we show the entanglement probabilities arising from the intersection and union of classes I and II, while in Fig.~\ref{fig:UnionIntersectionRatio}, we show the ratio of the union curve to the intersection curve. (We were able to obtain a formula for the former curve--but quite large in nature.)

To conclude this section, let us note that for the  witness $W_4$ specified by \cite[eq. (8.4)]{chruscinski2014entanglement}, we obtained an entanglement (without PPT-requirement) probability of $\frac{8}{25} \approx 0.32$ for the  $d=4$ Hiesmayr-L{\"o}ffler states, but no (with PPT-requirement) bound-entanglement.
Further, for the witness $W_4^{spin}$ specified by \cite[eq. (8.26)]{chruscinski2014entanglement}, no entanglement at all was detected.
\subsubsection{JBA witnesses} \label{IranianTwoququarts}
Analogously to our analyses in sec.~\ref{IranianTwoQutrits}, we constructed for the $d=4$ Hiesmayr-L{\"o}ffler model, the entanglement constraints for the pair of Jafarizadeh-Behzadi-Akbari witnesses $\mathcal{W}_{\alpha}$ and $\mathcal{W'}_{\alpha}$ given in \cite[eqs.(38) and (39)]{jafarizadeh2009entanglement}.
These took the forms,
\begin{equation}
 3 Q_4<\frac{(4 \alpha -1) \left(Q_1-4 Q_2\right)}{4 \alpha },
\end{equation}
and
\begin{equation}
3 \alpha  Q_4<\alpha  (4 \alpha -1) \left(2 Q_1+4 Q_2+4 Q_3-1\right),    
\end{equation}
respectively.

\begin{figure}
    \centering
    \includegraphics{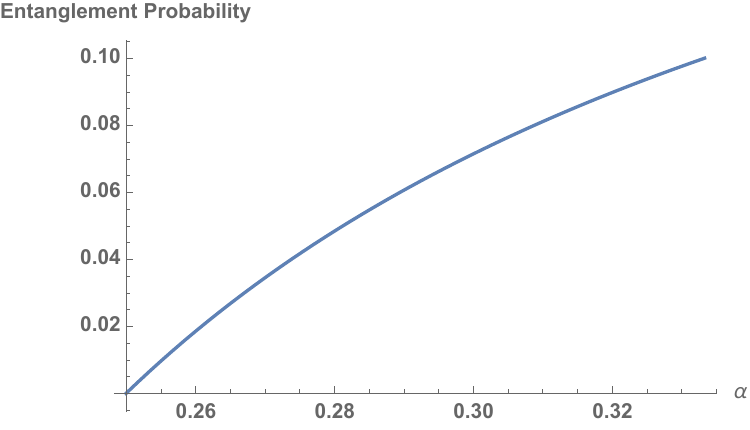}
    \caption{Two-ququart entanglement probability for the $d=4$ Hiesmayr-L{\"o}ffler model, $\alpha \in [\frac{1}{4},\frac{1}{3}]$, given by  (\ref{Firstentanglementprob})}
    \label{fig:IranianTwoququartsEntanglement}
\end{figure}

\begin{figure}
    \centering
    \includegraphics{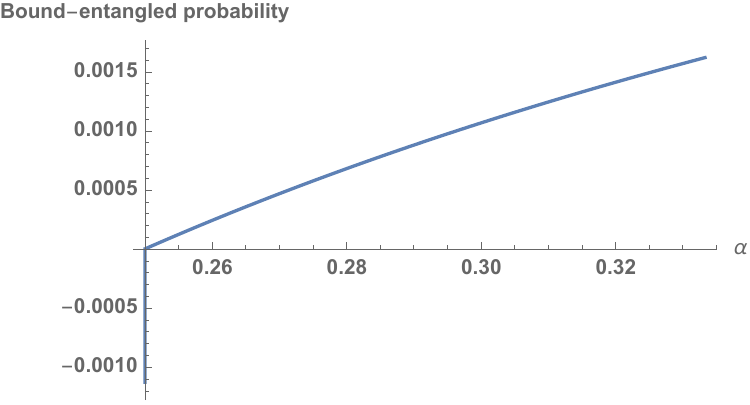}
    \caption{Two-ququart bound-entangled probability function (\ref{JBAbound}) for the $d=4$ Hiesmayr-L{\"o}ffler model, using either the JBA witness $\mathcal{W}_{\alpha}$ or $\mathcal{W'}_{\alpha}$, over $\alpha \in [\frac{1}{4},\frac{1}{3}]$.}
    \label{fig:IranianBoundTwoququarts}
\end{figure}

The entanglement probability based on either $\mathcal{W}_{\alpha}$ or  $\mathcal{W'}_{\alpha}$ took the form
\begin{equation} \label{Firstentanglementprob}
\frac{1-4 \alpha }{2-16 \alpha }  
\end{equation}
In Fig.~\ref{fig:IranianTwoququartsEntanglement},
we show this probability curve.
The (intersection) probability that both entanglement constraints are satisfied is $\frac{1-4 \alpha }{3(8 \alpha-1) }  $, while the (union) probability that at least one of the two is satisfied is $\frac{2-8 \alpha }{3-24 \alpha }$.
\paragraph{\underline{Third one-parameter family of bound-entangled probabilities.}} \label{thirdoneparameter}
Now, in Fig.~\ref{fig:IranianBoundTwoququarts}, we show the bound-entangled probability based on either the two-ququart JBA witnesses $\mathcal{W}_{\alpha}$ or  $\mathcal{W'}_{\alpha}$. This is given by 
\begin{equation} \label{JBAbound}
   \frac{A+B+C}{32 (4 \alpha -1)^{5/2} (20 \alpha +1)^{7/2}}, 
\end{equation}
where
\begin{displaymath}
 A=(-2 \alpha -1) (32 \alpha +1) (16 (1-5 \alpha ) \alpha +1)^2 \sqrt{\frac{6}{4 \alpha -1}+5},
\end{displaymath}
\begin{displaymath}
B= 27 \alpha  (8 \alpha +1)^3 (16 \alpha  (10 \alpha +1)+1) \cosh ^{-1}\left(\frac{12 \alpha }{8 \alpha +1}\right)
\end{displaymath}
and 
\begin{displaymath}
 C=-3 (8 \alpha +1)^4 (2 \alpha  (80 \alpha +23)+1) \text{sech}^{-1}\left(\sqrt{\frac{16 \alpha +2}{20 \alpha +1}}\right).
\end{displaymath}
 At $\alpha=\frac{3}{8} \approx 0.375$ (outside $[\frac{1}{4},\frac{1}{3}]$), the value is identical to the bound-entangled probability $\frac{24 \text{csch}^{-1}\left(\frac{8}{\sqrt{17}}\right)}{17 \sqrt{17}}-\frac{91}{544} \approx 0.00218722$, reported above in eq. (\ref{secondDariuz}), with respect to the indicated modification of the  entanglement witness $W_{2ququarts}$  (\ref{DariuzMatrix}).
 For given $\alpha \in  [\frac{1}{4},\frac{1}{3}]$, the two witnesses $\mathcal{W}_{\alpha}$ and $\mathcal{W'}_{\alpha}$ appear to form disjoint bound entanglement islands (cf. Fig.~\ref{fig:Void}).
 
\begin{figure}
     \centering
     \includegraphics{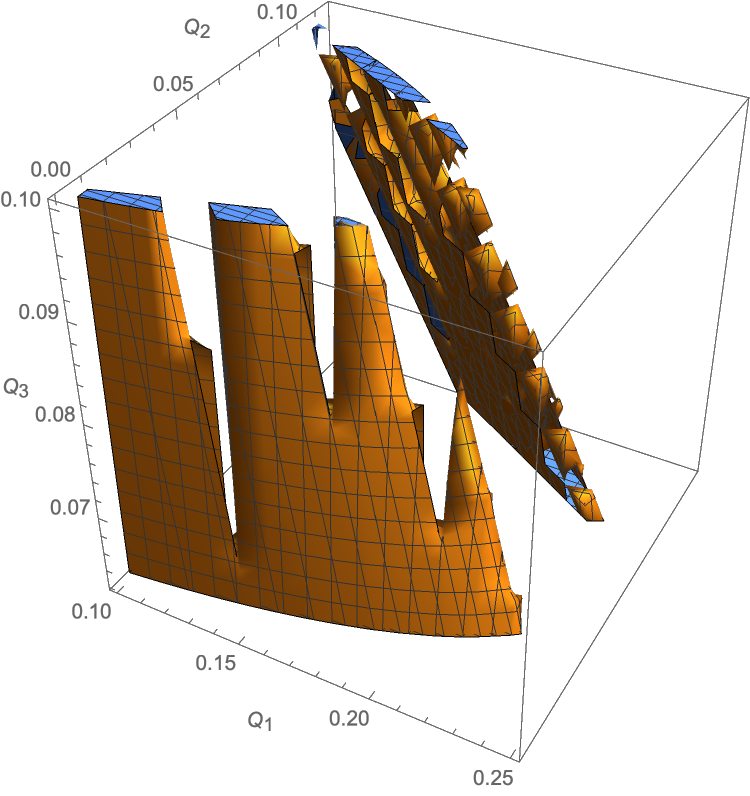}
     \caption{The anterior "jagged island" represents the  bound entanglement for the $d=4$ (two-ququart) Hiesmayr-L{\"o}ffler magic simplex revealed by the Jafarizadeh-Behzadi-Akbari witness $\mathcal{W}_{\alpha}$ and the posterior one, that shown by $\mathcal{W'}_{\alpha}$ . The parameter $\alpha$ has been set to $\frac{7}{24}$ and $Q_4$ to $\frac{1}{384}$.}
     \label{fig:IranianBoundEntanglement}
 \end{figure}

\begin{figure}
     \centering
     \includegraphics{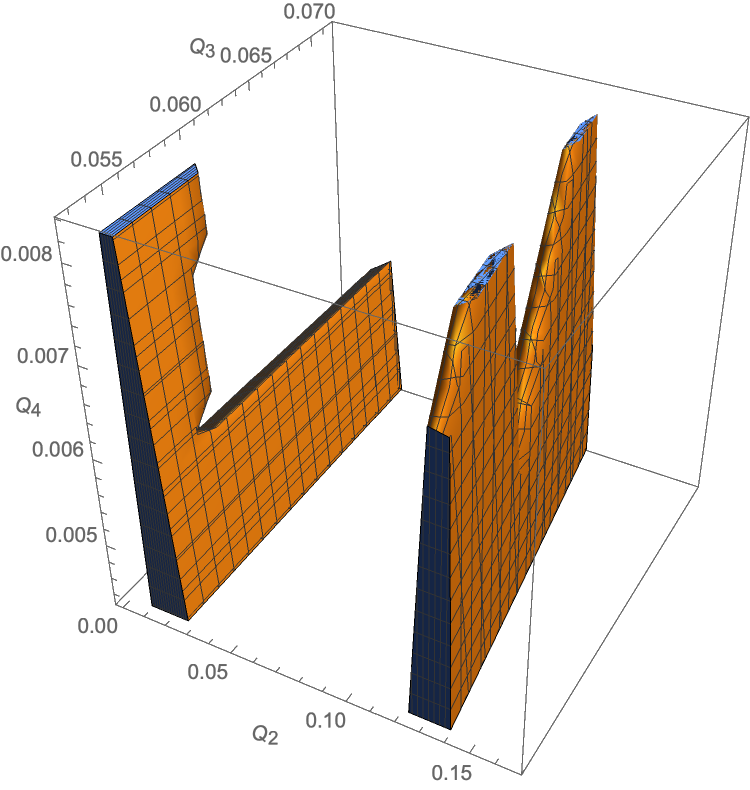}
     \caption{The left-positioned island represents the  bound entanglement based on the Jafarizadeh-Behzadi-Akbari witness $\mathcal{W}_{\alpha}$ and the right-side one that based on $\mathcal{W'}_{\alpha}$ for the $d=4$ (two-ququart) Hiesmayr-L{\"o}ffler magic simplex. The parameter $\alpha$ has been set to $\frac{1}{3}$ and $Q_1$ to $\frac{1}{6}$--so as to yield a three-dimensional image.}
     \label{fig:IranianBoundEntanglement2}
 \end{figure}

 In Fig.~\ref{fig:IranianBoundEntanglement} we show such a pair of islands for $\alpha = \frac{7}{24}$ and $Q_4 = \frac{1}{384}$,
 while in Fig.~\ref{fig:IranianBoundEntanglement2} we show such a pair of islands for $\alpha = \frac{1}{3}$ and $Q_1 = \frac{1}{6}$.
 
 The formula (\ref{JBAbound}) was derived assuming $\alpha \in [\frac{1}{4},\frac{1}{3}]$--as seemed suggested in \cite{jafarizadeh2009entanglement} (the authors claiming that the witnesses are non-decomposable [nd] there). Nevertheless, it appears to hold--as for $\alpha=\frac{3}{8}$--as well for $\alpha> \frac{1}{3}$. In the limit $\alpha \rightarrow \infty$, it attains the value
\begin{equation} \label{achievable?}
-\frac{5 \sqrt{5}+\log (8)-27 \log \left(3+\sqrt{5}\right)+24 \left(\log
   (2)+\text{csch}^{-1}(2)\right)}{200 \sqrt{5}} \approx    0.00728067.
\end{equation}
We are not aware of whether or not this is an achievable (or near-achievable) bound-entanglement probability measurement. But, for $\alpha =\frac{1}{3}$, the formula (\ref{JBAbound})  has the considerably lesser value $\frac{-92575+\frac{7798329 \cosh
   ^{-1}\left(\frac{12}{11}\right)}{\sqrt{23}}-\frac{13484361 \cosh
   ^{-1}\left(\sqrt{\frac{23}{22}}\right)}{\sqrt{23}}}{389344} \approx 0.00162026$.
\section{Realignment analyses of Hiesmayr-L{\"o}ffler magic simplices}
\subsection{Two-qutrit ($d=3$) case}
\begin{figure}
    \centering
    \includegraphics{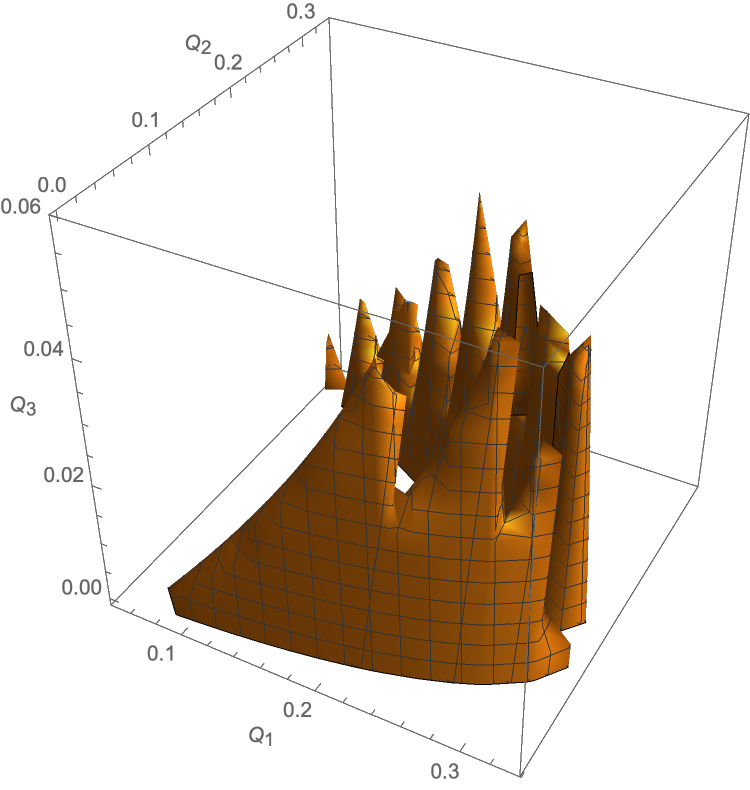}
    \caption{Islands of bound entanglement for $d=3$ Hiesmayr-L{\"o}ffler magic simplex model obtained on the basis of the realignment criterion for entanglement}
    \label{fig:RealignmentIsands}
\end{figure}

\begin{figure}
    \centering
    \includegraphics{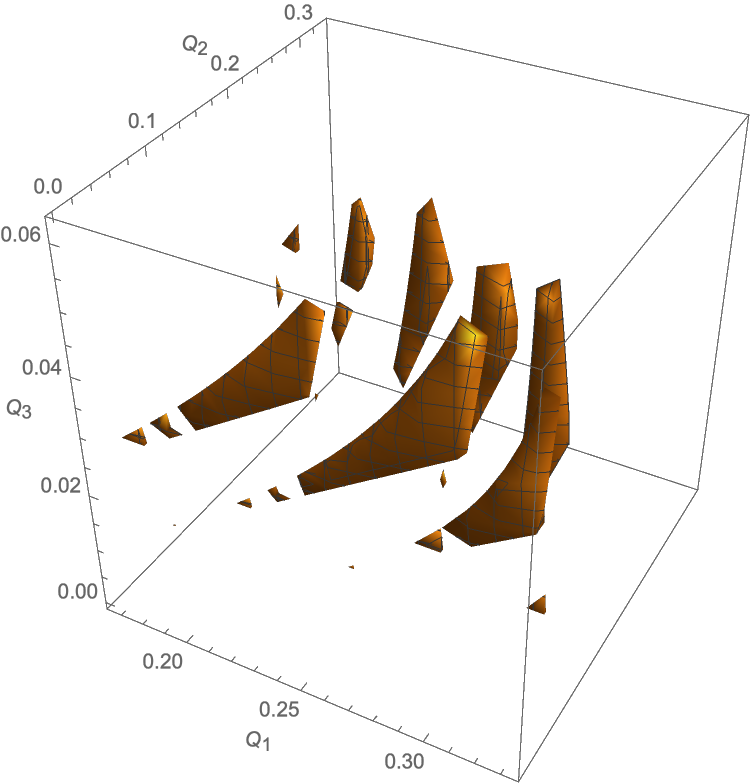}
    \caption{That part-of probability $\frac{1}{27} (10-9 \log (3)) \approx 0.00416627$--of the total realignment criterion bound entanglement (Fig.~\ref{fig:RealignmentIsands}), not found by either the Choi or  MUB witnesses.}
    \label{fig:ExcludedRegions}
\end{figure}

Application of the realignment (CCNR) test for entanglement \cite{chen2002matrix,shang2018enhanced} yielded an entanglement probability  of $\frac{1}{81} \left(27+\sqrt{3} \log \left(97+56 \sqrt{3}\right)\right) \approx 0.445977$ and an exact bound-entangled probability of $\frac{2}{81} \left(4 \sqrt{3} \pi -21\right) \approx 0.0189035$, considerably larger than the 0.00736862 and 0.00325613 reported above. The realignment constraint that, if satisfied, ensures entanglement is 
\begin{equation}
 \frac{2}{3} \sqrt{-9 Q_2-6 Q_3+3 \left(Q_1^2+\left(3 Q_2+4 Q_3-1\right) Q_1+9 Q_2^2+4
   Q_3^2+6 Q_2 Q_3\right)+1}+
\end{equation}
\begin{displaymath}
 1/3+ |Q_1-Q_3| >1.
\end{displaymath}
The bound-entanglement islands obtained by enforcing this constraint are displayed in Fig.~\ref{fig:RealignmentIsands}. (It is interesting to note that the region of {\it free} entanglement also, in fact, separates into two--not as severely jagged--islands of its own.) These realignment islands completely contain the corresponding Choi and MUB islands, with an additional probability of $\frac{1}{27} (10-9 \log (3)) \approx 0.00416627$.
This (highly fragmented) domain of additional bound-entangled probability found through use of the realignment criterion, but by neither the Choi nor the MUB witnesses is shown in Fig.~\ref{fig:ExcludedRegions}.

We also implemented the  entanglement detection criterion (ESIC) based on symmetric informationally complete positive operator-valued measures (SIC POVMs) \cite{shang2018enhanced}. We found it to be equivalent to the realignment (CCNR) test.

Motivated by the recent preprint, ``Bound entangled states fit for robust experimental verification'' \cite{sentis2018bound}, we found that the CCNR reached a maximum of $\frac{7}{6}$ at $Q_1=\frac{1}{4}, Q_2=\frac{1}{24} (3 -\sqrt{5}), Q_3=0$. (In the original Hiesmayr-L{\"o}ffler coordinates, this converts to $q_1 = \frac{5}{24} \left(\sqrt{5}-3\right),q_2=
   -1-\frac{\sqrt{5}}{3},q_3=-\frac{\sqrt{5}}{4}$.) It would be interesting to further analyze this result in the framework of \cite{sentis2018bound}, to
ascertain how robust--in terms of their criteria--this state is for experimental verification.
The (rank-7) density matrix in question takes the form (zeros being represented by dots)
\begin{equation} \label{Sentis}
\left(\mathcode`0=\cdot
\begin{array}{ *{3}{c} | *{3}{c} | *{3}{c} }
 \frac{1}{12} & 0 & 0 & 0 & \frac{1}{12} & 0 & 0 & 0 & \frac{1}{12} \\
 0 & \frac{1}{24} \left(3-\sqrt{5}\right) & 0 & 0 & 0 & 0 & 0 & 0 & 0 \\
 0 & 0 & \frac{1}{24} \left(3+\sqrt{5}\right) & 0 & 0 & 0 & 0 & 0 & 0 \\
 0 & 0 & 0 & \frac{1}{24} \left(3+\sqrt{5}\right) & 0 & 0 & 0 & 0 & 0 \\
 \frac{1}{12} & 0 & 0 & 0 & \frac{1}{12} & 0 & 0 & 0 & \frac{1}{12} \\
 0 & 0 & 0 & 0 & 0 & \frac{1}{24} \left(3-\sqrt{5}\right) & 0 & 0 & 0 \\
 0 & 0 & 0 & 0 & 0 & 0 & \frac{1}{24} \left(3-\sqrt{5}\right) & 0 & 0 \\
 0 & 0 & 0 & 0 & 0 & 0 & 0 & \frac{1}{24} \left(3+\sqrt{5}\right) & 0 \\
 \frac{1}{12} & 0 & 0 & 0 & \frac{1}{12} & 0 & 0 & 0 & \frac{1}{12} \\
\end{array}
\right).
\end{equation}
However, undermining the initially presumed robustness of this state is the observation that three of the eigenvalues of its partial transpose are zero, so any perturbation of the state might lead to negative eigenvalues of the partial transpose, and thus its departure from the PPT domain.
\subsection{Two-ququart ($d=4$) case}
Here, we observed  free entanglement and bound-entangled probability CCNR-based estimates of 0.4509440211445637 and 0.01265489845176, respectively.
\section{Generalized Horodecki state analyses} \label{GeneralizedHorodecki}
At the conclusion of their paper \cite{jafarizadeh2009entanglement}, Jafarizadeh,  Behzadi, and Akbari consider generalized Horodecki states of the form $\rho_i =\Sigma_{i=1}^n a_i O_i$, where the $O_i$'s are unit-trace orthogonal operators. These were used as a basis for constructing EWs, in particular, the pair of witnesses $\mathcal{W}_{\alpha}$ and $\mathcal{W'}_{\alpha}$ we have employed above in certain two-qutrit (sec.~\ref{IranianTwoQutrits}) and two-ququart (sec.~\ref{IranianTwoququarts}) analyses. (Let us also note that Chru{\'s}ci{\'n}ski and  Rutkowski have provided a multi-parameter family of 2-qudit PPT entangled states which generalize the celebrated Horodecki-state \cite{chruscinski2011family}.)
\subsection{Two-qutrit case} \label{trits}
Following the suggested approach of JBA for obtaining generalized Horodecki states, in the two-qutrit instance, we have found a PPT-probability of $\frac{1}{27} \left(4 \sqrt{3} \pi -9\right) \approx 0.4278$. The entanglement probability with either of the two  witnesses is $\frac{1}{2}$. The bound-entangled probability for either witness is  $\frac{1}{27} \left(2 \sqrt{3} \pi -9\right) \approx 0.0697322$ for $\alpha \in [\frac{1}{3},\frac{2}{3}]$.
\subsection{Two-ququart case} \label{quarts}
In the two-ququart generalized Horodecki case, the PPT-probability is 
$\frac{1}{3} \approx 0.3333$. The entanglement probability with either witness is $\frac{1}{2}$--again, independently of $\alpha$. The bound-entangled probability with either witness is  $\frac{1}{24} \approx 0.0416667$ for $\alpha \in [\frac{1}{4},\frac{1}{3}]$. 

\begin{figure}
     \centering
     \includegraphics{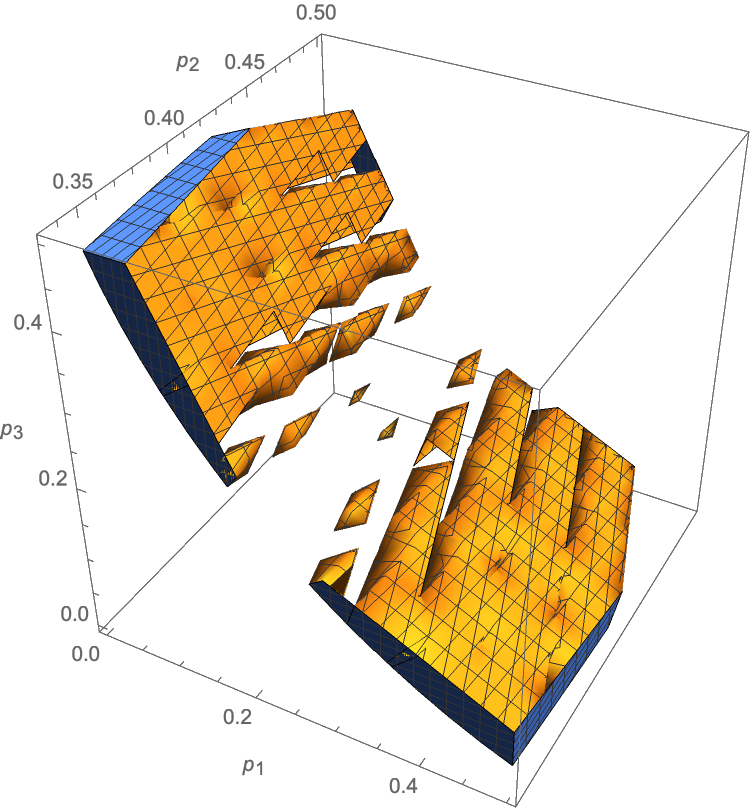}
     \caption{The upper ``jagged island/archipelago" represents the  bound entanglement based on the Jafarizadeh-Behzadi-Akbari witness $\mathcal{W}_{\alpha}$ and the lower one that based on $\mathcal{W'}_{\alpha}$ for the  two-ququart generalized Horodecki states.}
     \label{fig:CraggyHorodecki}
 \end{figure}

In Fig.~\ref{fig:CraggyHorodecki} we show a pair of jagged bound-entanglement islands. Six times the displayed volume for each island gives the bound-entanglement probability of $\frac{1}{24}$. (These plots are independent of $\alpha$.)
\subsection{Two-ququint case}
For the $5 \times 5$ two-ququint ($25 \times 25$ density matrices) scenario, the entanglement probability is simply again $\frac{1}{2}$, independently of $\alpha$, for the two witnesses.  (This case is not explicitly discussed in \cite{jafarizadeh2009entanglement}, and we have followed their discussions of the previous lower-dimensional instances.) The PPT-probability is approximately 0.33734924124312192527. The bound-entangled probability is $\approx 0.0370662$ for $\alpha > \frac{1}{5}$. 

It would be of interest to investigate the relations between the JBA generalized Horodecki states we have analyzed and those put forth by  Chru{\'s}ci{\'n}ski and  Rutkowski  \cite{chruscinski2011family}.
\section{Full-dimensional magic simplices numerical analyses}
\subsection{Two-qutrits}
Moving on from the $d=3$ Hiesmayr-L{\"o}ffler model, we examined the original 8-dimensional ``magical" simplex of Bell states of bipartite qutrits  studied by Baumgartner, Hiesmayr and Narnhofer \cite{baumgartner2006state,baumgartner2008geometry}.

\begin{figure}
    \centering
    \includegraphics{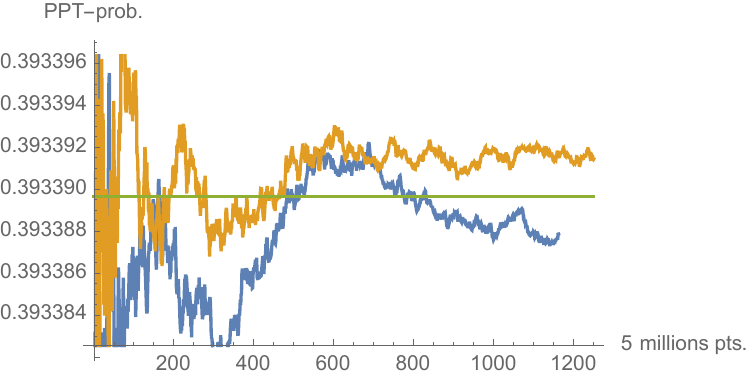}
    \caption{A pair of  (quasi-random) PPT-probability estimates for the 8-dimensional magical simplex of Bell states of bipartite qutrits. The last recorded estimates are   0.39338785 (Roberts parameter $\alpha_0 = \frac{1}{2}$) and 0.39339143 ($\alpha_0 = 0$). The plotted value $\frac{7 \pi}{25 \sqrt{5}} \approx 0.3933896249$ provides a close fit, with the weighted average of the two estimates being 0.3933897072. We note that the Hiesmayr-L{\"o}ffler $d=3$ counterpart is $\frac{8 \pi }{27 \sqrt{3}} \approx 0.537422$.}
    \label{fig:Magic8D}
\end{figure}

In Fig.~\ref{fig:Magic8D}, we plot--using the interesting (golden-ratio-related) ``quasi-random" procedure (we have been recently employing \cite{slater2019quasirandom}) of Martin Roberts \cite{Roberts32D,Roberts}--two sets of estimates of the associated Hilbert-Schmidt PPT-probability. One (symmetrically) fixes the Roberts parameter $\alpha_0$ at $\frac{1}{2}$ (which may lead to superior estimates), and the other at zero. Large numbers of realizations--5,820,000,000 realizations in the former case, and 6,260,000,000 in the latter--were used.  The two estimates obtained were 0.39338785 and 0.39339143, respectively. A well-fitting conjecture for the underlying exact value--of a seemingly similar nature to the $d=3$ Hiesmayr-L{\"o}ffler counterpart of $\frac{8 \pi }{27 \sqrt{3}}$--is $\frac{7 \pi}{25 \sqrt{5}} \approx 0.3933896249$. (In particular, the average of the two estimates appears to very strongly converge in this direction--with the last recorded average agreeing with the conjecture to six decimal places.)

In a supplementary analysis to these two, now using a Roberts parameter of $\alpha_0=\frac{1}{3}$, together with 280,000,000 million realizations, we obtained an estimate of 0.393381, plus, additionally,  an estimate of 0.00011335 for the associated bound-entangled probability based on the Hiesmayr-L{\"o}ffler mutually-unbiased-bases criterion (\ref{twoparameters}) of \begin{equation}
I_4=3 c[0, 0] + c[0, 1] + 2 c[0, 2] + c[1, 1] + 2 c[1, 2] + c[2, 1] + 
 2 c[2, 2]>2.
\end{equation}
Let us here recall that for the Hiesmayr-L{\"o}ffler $d=3$ counterpart, the bound-entangled probability was found to be considerably greater, that is,
$-\frac{4}{9}+\frac{4 \pi }{27 \sqrt{3}}+\frac{\log (3)}{6} \approx 0.00736862$, than  0.000113354.
\subsection{Two-ququarts}
For the 15-dimensional magic simplex of Bell states of bipartite ququarts, two Hilbert-Schmidt PPT-probability quasi-random-based estimates, employing approximately one hundred seventy million realizations each, were 0.115717 ($\alpha_0=\frac{1}{2}$) and 0.115778 ($\alpha_0=0$). The Hiesmayr-L{\"o}ffler counterpart is $\frac{1}{2}+\frac{\log \left(2-\sqrt{3}\right)}{8 \sqrt{3}} \approx 0.404957$.

\begin{figure}
    \centering
    \includegraphics{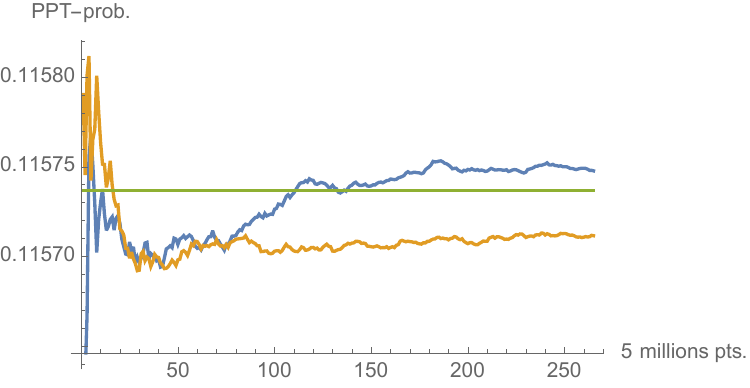}
    \caption{A pair of  (quasi-random) PPT-probability estimates for the 15-dimensional magical simplex of Bell states of bipartite ququarts. The last recorded estimates are 0.1157477026  (Roberts parameter $\alpha_0 = \frac{1}{4}$) and 0.1157115381 ($\alpha_0 = \frac{3}{4}$). Also shown is a conjectured value of $\frac{1}{8}+\frac{\log \left(3-\sqrt{5}\right)}{13 \sqrt{5}} \approx 0.115737$. We note that the Hiesmayr-L{\"o}ffler $d=4$ counterpart is $\frac{1}{2}+\frac{\log \left(2-\sqrt{3}\right)}{8 \sqrt{3}} \approx 0.404957$.}
    \label{fig:Magic15DPPT}
\end{figure}

At this point--due to certain technical issues--we undertook a parallel set of analyses (however, shifting the Roberts parameters from 0 and $\frac{1}{2}$ to $\frac{1}{4}$ and $\frac{3}{4}$). Additionally, now we recorded whether or not the realignment test \cite{chen2002matrix} for entanglement was passed. In Fig.~\ref{fig:Magic15DPPT}, we plot the two sets of estimates of the associated Hilbert-Schmidt PPT-probability. An interesting candidate value is $\frac{1}{8}+\frac{\log \left(3-\sqrt{5}\right)}{13 \sqrt{5}} \approx 0.115737$.

\begin{figure}
    \centering
    \includegraphics{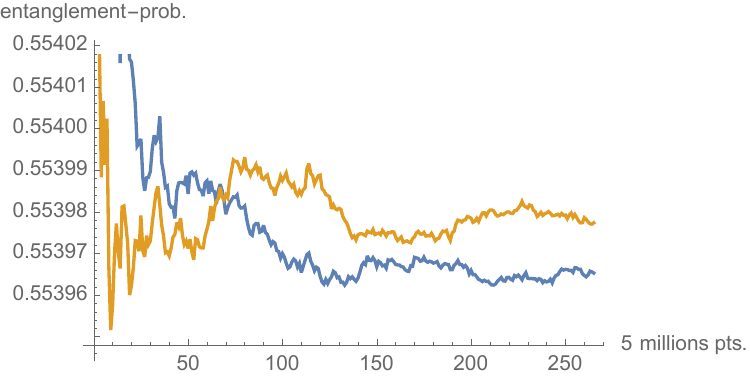}
    \caption{A pair of  (quasi-random) entanglement-probability estimates--based on the realignment criterion--for the 15-dimensional magical simplex of Bell states of bipartite ququarts. The last recorded estimates are   0.5539652981 (Roberts parameter $\alpha_0 = \frac{1}{4}$) and 0.5539773751 ($\alpha_0 = \frac{3}{4}$).}
    \label{fig:Magic15DEntanglement}
\end{figure}

Further, in Fig.~\ref{fig:Magic15DEntanglement}, we plot the two sets of estimates of the associated Hilbert-Schmidt entanglement probability based on the realignment criterion.

\begin{figure}
    \centering
    \includegraphics{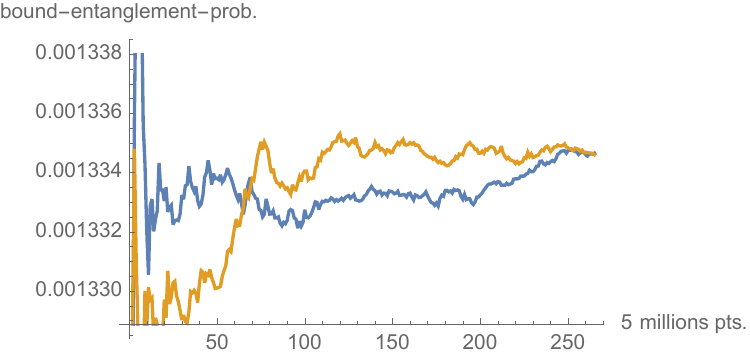}
    \caption{A pair of  (quasi-random) bound-entanglement-probability estimates--based on the realignment criterion--for the 15-dimensional magical simplex of Bell states of bipartite ququarts. The last recorded estimates are 0.001334655094 (Roberts parameter $\alpha_0 = \frac{1}{4}$) and 0.001334600000 ($\alpha_0 = \frac{3}{4}$).  A conjecture of $\frac{1}{750} \approx 0.00133333$ can be advanced.}
    \label{fig:Magic15DBound}
\end{figure}

Also, in Fig.~\ref{fig:Magic15DBound}, we display the pairs of estimates of the Hilbert-Schmidt bound-entangled probability based on the realignment criterion. A conjecture of $\frac{1}{750} \approx 0.00133333$ can be advanced, in this regard.
\section{Further discussion}
It has been established that for the two-rebit, rebit-retrit and two-retrit $X$-states (the density matrices for which, by definition,  have their only nonzero entries along their diagonal and anti-diagonal \cite{Rau:2018:CQD:3268457.3268690}), the Hilbert-Schmidt separability/PPT probabilities are all equal to $\frac{16}{3 \pi^2}\approx 0.54038$ for the two-rebit, rebit-retrit and two-retrit $X$-states (cf. \cite{dunkl2015separability}). Numerical and exact analyses of ours strongly indicated that among the ($9 \times 9$) two-retrit PPT-states, none is bound entangled in terms of the Hiesmayr-L{\"o}ffler MUB $I_d>2$ criterion. But the question of whether there are bound-entangled $X$-states should be addressed more thoroughly  than so far has been done.

There certainly are many more directions in which efforts to determine  probabilities of bound-entangled states can be directed (cf. \cite[sec.IV.C]{zyczkowski1999volume} for a $2 \times 4$ density matrix analysis, and \cite{zhou2018quantum} for multipartitite issues). 

Much research has been devoted to the determination of Hilbert-Schmidt (and other--Bures, monotone) separability and PPT-probabilities \cite{slater2017master,slater2018extensions,slater2019quasirandom} (and references therein), but considerably less so, it would seem, as we have attempted here, to the bound-entangled situation. (Perhaps we can regard the Horodecki-state bound-entangled probability of $\frac{1}{5}$ noted above, as the initial result in this area of research.) But of interest in these respects, is the paper \cite{bae2009detection}, in which there was derived ``an explicit analytic estimate for the entanglement of a large class of bipartite quantum states, which extends into bound entanglement regions''.

As to the full 35-dimensional set of two-qutrit states itself, evidence has been presented indicating that the associated Hilbert-Schmidt PPT-probability--on the order of 0.0001027 \cite[Fig. 9]{slater2019quasirandom}--is constant over the Casimir invariants of their qutrit subsystems \cite[sec.III.A]{slater2016invariance}. (Also, a ``repulsion'' effect for the Casimir invariants has been observed in the two-qubit case, where the invariants are the Bloch radii of the individual qubits \cite{slater2016two}. The two-qutrit case was also examined there.)

In an auxiliary full 63-dimensional $2 \times 4$ qubit-qudit analysis, based on 1,200 million iterations, 
use of the realignment criterion \cite{chen2002matrix} yielded an 
estimate of 0.00023410917 for the bound-entangled probability  and 0.94234319 (conjecturally, $\frac{589}{625}=\frac{17 \cdot 31}{5^4} \approx 0.94234616$) for for the entanglement probability, in general. In that (quasirandom Hilbert-Schmidt) analysis, we were not able to detect any finite probability at all of genuinely tripartite entanglement using the Greenberger-Horne-Zeilinger test set out in Example 3 in \cite{bae2018entanglement}. (However, in a parallel full 80-dimensional two-qutrit study, the realignment test for entanglement was not passed by any randomly generated states (cf.\cite{gabdulin2019investigating}).)

As a final remark, let us conjecture that the "nonsmooth/jagged" nature of the boundaries of regions of bound entanglement reported above (Figs~\ref{fig:Void}, \ref{fig:IranianBoundEntanglementd3}, \ref{fig:IranianBoundEntanglement}, \ref{fig:IranianBoundEntanglement2} and \ref{fig:CraggyHorodecki}), will, in some sense--remaining to be made precise--diminishes with increasing dimensions of, say, bipartite systems (cf. \cite{beigi2010approximating}).
\section{Compliance with Ethical Standards}
The author asserts that he has no conflicts of interest
or potential such conflicts. The research reported did not involve human participants and/or animals.
\appendix
\section{{Generalized Horodecki-Werner States}}
\subsection{Two-qutrit case}
Let us consider the states that are composed of equally-weighted two-qutrit generalized Horodecki ones (in the sense discussed [sec.~\ref{trits}]) and the fully mixed two-qutrit state.

Then the PPT-probability for this set is
\begin{equation}
 \frac{1}{81} \left(-66+21 \sqrt{17}+50 \sqrt{3} \sin ^{-1}\left(\frac{1}{50}
   \sqrt{\frac{3}{2} \left(417-7 \sqrt{17}\right)}\right)\right) \approx    0.792568.
\end{equation}
The entanglement probability for either Jafarizadeh-Behzadi-Akbari (JBA) witness $\mathcal{W}_{\alpha}$ or $\mathcal{W'}_{\alpha}$ is given by
\begin{equation} \label{Equal3by3}
 \begin{cases}
 \frac{(7-15 \alpha )^2}{162 (1-3 \alpha )^2} & 15 \alpha >7\lor \alpha <0 \\
 \frac{3 \alpha  (411 \alpha -254)+113}{162 (1-3 \alpha )^2} & 0<\alpha \leq \frac{1}{6}
   \\
 \frac{(11-39 \alpha )^2}{162 (1-3 \alpha )^2} & \frac{1}{6}<\alpha <\frac{11}{39}
\end{cases} .
\end{equation}
The bound-entangled probability for either witness is
\begin{equation}
 \frac{3 \alpha  (411 \alpha -254)+113}{162 (1-3 \alpha )^2},  
\end{equation}
for $ \alpha \in [\frac{1}{309} \left(71-6 \sqrt{17}\right) \approx 0.149713,\frac{1}{6}]$
and
\begin{equation}
 \frac{(11-39 \alpha )^2}{162 (1-3 \alpha )^2} , 
\end{equation}
for $ \alpha \in [\frac{1}{6}, \frac{11}{39}]   $.
(In these last two intervals, all the entanglement probability is bound--as can be seen from (\ref{Equal3by3}).)
\subsection{Two-ququart case}
\begin{figure}
    \centering
    \includegraphics{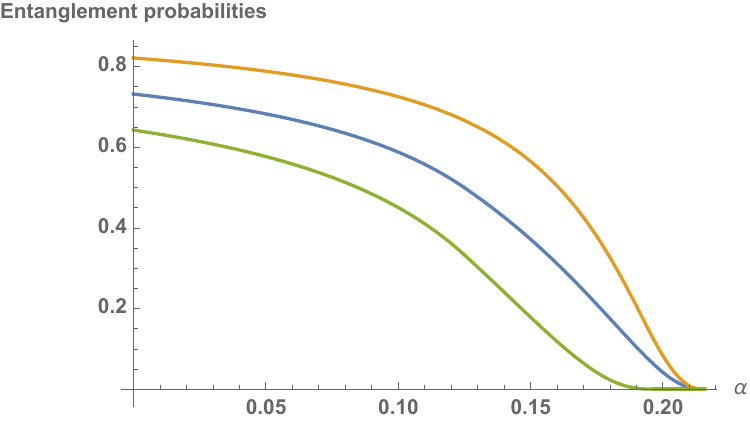}
    \caption{The equally-weighted (Horodecki-Werner) two-ququart entanglement probability function 
    (\ref{WernerEntanglement}) for either JBA witness $\mathcal{W}_{\alpha}$ or $\mathcal{W'}_{\alpha}$, along with the dominant union and subordinate intersection curves, obtained by application of the two witnesses. The interval employed is $\alpha \in [0,\frac{19}{88}]$.}
    \label{fig:Werner4by4}
\end{figure}

\begin{figure}
    \centering
    \includegraphics{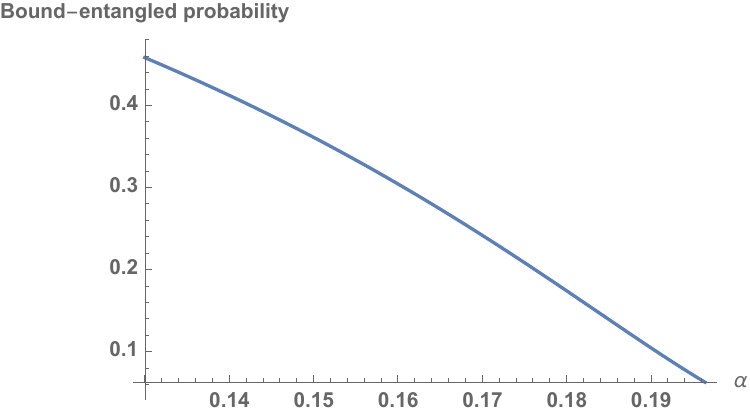}
    \caption{Bound-entangled probability function (\ref{Werner4by4BoundEquation}) for the equally-weighted Horodecki-Werner two-ququart scenario--restricted to the interval $\alpha \in  [1/872 (161 - 18 \sqrt{7}) \approx 0.130019,\frac{11}{56} \approx 0.196429]$.}
    \label{fig:Werner4by4Bound}
\end{figure}

Now, for the class of equally-weighted generalized Horodecki two-ququart states (in the above JBA sense again [sec.~\ref{quarts}]) and the fully mixed two-ququart state, the PPT-probability is 
\begin{equation}
    \frac{437}{192}-\frac{7 \sqrt{7}}{12} \approx 0.732687.
\end{equation}
The entanglement probability for either JBA witness is 
\begin{equation} \label{WernerEntanglement}
\begin{cases}
 \frac{(40 \alpha -13)^3}{8192 (4 \alpha -1)^3} & \alpha >\frac{13}{40}\lor \alpha <0 \\
 \frac{(88 \alpha -19)^3}{8192 (4 \alpha -1)^3} & \frac{1}{8}<\alpha <\frac{19}{88} \\
 \frac{460288 \alpha ^3-330816 \alpha ^2+78024 \alpha -5995}{8192 (4 \alpha -1)^3} &
   0<\alpha <\frac{1}{8}
\end{cases}.   
\end{equation}
In Fig.~\ref{fig:Werner4by4}, we show this function over the interval $\alpha \in \frac{1}{8},\frac{19}{48}]$ along with the entanglement probabilities based on the union and intersection of application of the two witnesses.
In a somewhat indirect fashion, We have been able to establish--through examination of numerous specific values of $\alpha$ and certain auxiliary analyses--that the bound-entangled probability function for either witness takes the form (Fig.~\ref{fig:Werner4by4Bound})
\begin{equation} \label{Werner4by4BoundEquation}
 \frac{8 \alpha  (8 \alpha  (14600 \alpha -9537)+16599)-9623}{12288 (4 \alpha -1)^3} ,  
\end{equation}
over  $\alpha \in  [1/872 (161 - 18 \sqrt{7}) \approx 0.130019,\frac{11}{56} \approx 0.196429]$. The nature of the function outside this interval remains under examination.
\subsection{Two-ququints}
The entanglement probability for this still higher-dimensional equally-weighted case for either $\mathcal{W}_{\alpha}$ or $\mathcal{W'}_{\alpha}$ is 
\begin{equation}
\begin{cases}
  \frac{(85 \alpha -21)^4}{781250 (5 \alpha -1)^4} & \alpha >\frac{21}{85}\lor \alpha <0
   \\
 \frac{(165 \alpha -29)^4}{781250 (5 \alpha -1)^4} & \frac{1}{10}<\alpha <\frac{29}{165}
   \\
 \frac{436080625 \alpha ^4-339038500 \alpha ^3+98070150 \alpha ^2-12476260 \alpha
   +586769}{781250 (5 \alpha -1)^4} & 0<\alpha \leq \frac{1}{10}
\end{cases}   .
\end{equation}
The PPT-probability is $\approx 0.758301$. At $\frac{1}{6}$, the bound-entangled probability is, interestingly,  $\frac{6561}{781250} =\frac{3^8}{2 \cdot 5^8} \approx 0.00839808$. It jumps at $\alpha=\frac{1}{7}$ to 0.1646314041874492= 
0.0119320240085435+$\frac{8}{625}+\frac{2008781-157760 \sqrt{145}}{1953125}+\frac{72 \left(696
   \sqrt{145}-7925\right)}{390625}$.
   
\begin{figure}
    \centering
    \includegraphics{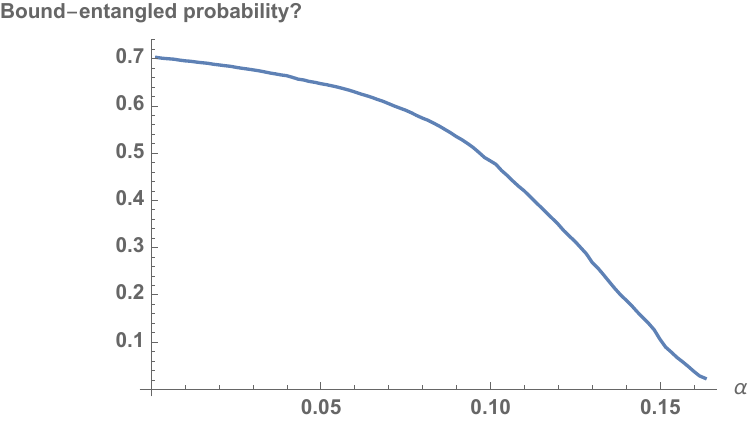}
    \caption{Numerically-derived plot for equally-weighted two-ququint scenario over $\alpha \in [0,\frac{1}{6}]$.}
    \label{fig:BoundEntanglementPlot}
\end{figure}

In Fig.~\ref{fig:BoundEntanglementPlot}, we show a numerically-derived plot of this function over $\alpha \in [0,\frac{1}{6}]$.
   
An important question that remains is whether outside the intervals employed by Jafarizadeh, Behzadi, and Akbari, $\mathcal{W}_{\alpha}$ and $\mathcal{W'}_{\alpha}$ can serve as entanglement witnesses--and, thus,  whether a number of probabilities presented above in this section that were termed ``entanglement" and ``bound-entangled" are. in fact, so.

\begin{acknowledgements}
This research was supported by the National Science Foundation under Grant No. NSF PHY-1748958. Nicolas Tessore greatly assisted in the carrying out of many of the calculations reported above. I also strongly thank Dariuz  Chru{\'s}ci{\'n}ski for sharing his entanglement-witness expertise.
\end{acknowledgements}

\bibliography{main}

\end{document}